\documentclass[twocolumn]{openjournal}

\usepackage{xcolor}
\usepackage{textgreek}
\usepackage[utf8]{inputenc}
\usepackage[english]{babel}
\usepackage{natbib}
\usepackage{color,colortbl}
\definecolor{linkcolor}{rgb}{0.0,0.3,0.5}
\usepackage{tensind}
\tensordelimiter{?}
\DeclareGraphicsExtensions{.bmp,.png,.jpg,.pdf}
\usepackage{verbatim}
\usepackage[normalem]{ulem}
\usepackage{orcidlink}
\usepackage{soul}
\usepackage{bm}

\graphicspath{ {./figs/} }

\usepackage{hyperref}
\hypersetup{
    unicode, 
    colorlinks=true,
    linkcolor=linkcolor,
    citecolor=linkcolor,
    filecolor=linkcolor,
    urlcolor=linkcolor,
}

\makeatletter 
  \patchcmd{\NAT@citex}
    {\@citea\NAT@hyper@{%
      \NAT@nmfmt{\NAT@nm}%
      \hyper@natlinkbreak{\NAT@aysep\NAT@spacechar}{\@citeb\@extra@b@citeb}%
      \NAT@date}}
    {\@citea\NAT@nmfmt{\NAT@nm}%
    \NAT@aysep\NAT@spacechar\NAT@hyper@{\NAT@date}}{}{}

  \patchcmd{\NAT@citex}
    {\@citea\NAT@hyper@{%
      \NAT@nmfmt{\NAT@nm}%
      \hyper@natlinkbreak{\NAT@spacechar\NAT@@open\if*#1*\else#1\NAT@spacechar\fi}%
        {\@citeb\@extra@b@citeb}%
      \NAT@date}}
    {\@citea\NAT@nmfmt{\NAT@nm}%
    \NAT@spacechar\NAT@@open\if*#1*\else#1\NAT@spacechar\fi\NAT@hyper@{\NAT@date}}
    {}{}
\makeatother

\newcommand\blfootnote[1]{%
  \begingroup
  \renewcommand\thefootnote{}\footnote{#1}%
  \addtocounter{footnote}{-1}%
  \endgroup
}

\begin{document}
\title{The birth of the intracluster medium: the evolution of multiphase gas and Lyman-$\bm{\alpha}$ haloes in a simulated $z\sim3$ protocluster}

\author{Jake S. Bennett$^{1,2,\star}$\orcidlink{0000-0002-8573-2993}}
\author{Aaron Smith$^3$\orcidlink{0000-0002-2838-9033}}
\author{Fabrizio Arrigoni-Battaia$^4$\orcidlink{0000-0002-4770-6137}}
\author{Debora Sijacki$^{5,6}$}
\author{\\Cassandra Lochhaas$^{1,7}$\orcidlink{0000-0003-1785-8022}}
\author{Lars Hernquist$^1$}

\affiliation{$^1$Center for Astrophysics $|$ Harvard \& Smithsonian, 60 Garden Street, Cambridge, MA 02138, USA \\ 
$^2$School of Physics \& Astronomy, University of Nottingham, University Park, Nottingham NG7 2RD, UK\\
$^3$Department of Physics, The University of Texas at Dallas, Richardson, Texas 75080, USA \\ 
$^4$Max-Planck-Institut für Astrophysik, Karl-Schwarzschild-Straße 1, 85748 Garching bei München, Germany \\
$^5$Institute of Astronomy, University of Cambridge, Madingley Road, Cambridge, CB3 0HA, UK \\
$^6$Kavli Institute for Cosmology, Cambridge, University of Cambridge, Madingley Road, Cambridge, CB3 0HA, UK\\
$^7$NASA Hubble Fellow}

\blfootnote{$^{\star}$\mbox{Email: \href{mailto:jake.bennett@cfa.harvard.edu}{jake.bennett@cfa.harvard.edu}}}

\begin{abstract}
Galactic haloes host a complex, multiphase circumgalactic medium (CGM), and at high redshift are fed by cold, filamentary inflows. In contrast, mature galaxy clusters are dominated by a hot, enriched, X-ray emitting intracluster medium (ICM), with cold gas largely confined to member galaxies. However, the transition between these regimes remains poorly constrained. We present a cosmological zoom-in simulation of a massive cluster progenitor evolved to $z=2.7$, with enhanced CGM resolution to better trace the accretion, mergers and feedback events that precede the birth of the ICM. We connect this evolution to mock Mg\,\textsc{ii}, C\,\textsc{ii}, O\,\textsc{vi} and O\,\textsc{vii} absorption, tracing low and high ionisation gas phases. We also study Lyman-$\alpha$ (Ly$\alpha$) and Balmer-$\alpha$ (H$\alpha$) haloes in emission, using radiative transfer in post-processing. Between $z\sim4.4$ and $2.7$, a major merger and AGN feedback drive an inside-out transformation, redistributing gas to larger radii and flattening density, temperature and metallicity profiles. Intermediate column Mg\,\textsc{ii} absorbers are rapidly destroyed, leaving a clumpier cold gas distribution associated with satellites, while gas is ionised beyond O\,\textsc{vii} as the inner halo enters the X-ray regime. An extended Ly$\alpha$ halo remains detectable even without AGN photoionisation, and evolves from filamentary to more spherical as inflowing gas is disrupted. Our fiducial model underpredicts observed central Ly$\alpha$ emission --- we likely require more efficient Ly$\alpha$ production in the nuclear region, either through more effective escape of stellar Ly$\alpha$ photons or through enhanced conversion of AGN-powered ionisation into Ly$\alpha$ emission. H$\alpha$ haloes are dimmer and smaller than Ly$\alpha$, but with \textit{JWST} may provide a complementary probe of the evolving CGM at this critical epoch. 
\end{abstract}

\begin{keywords}
    {galaxies: formation, galaxies: evolution, galaxies: clusters: intracluster medium, methods: numerical}
\end{keywords}

\maketitle

\section{Introduction}
\label{Section:Intro}

Abundant observational evidence now points to most galaxies being surrounded by a complex and strongly multiphase circumgalactic medium (CGM), with inflowing, outflowing, hot and cold material all coexisting and interacting \citep[see e.g.][]{Tumlinson2017,FaucherGiguereOh2023,GronkeSchneider2026}. At high redshift, this multiphase CGM often takes the form of cold, dense filaments penetrating through a hotter halo to feed early galaxy formation \citep{BirnboimDekel2003,Katz2003,Keres2005,Dekel2009,vandeVoort2011,vandeVoort2012,Bennett2020,Mandelker2020a,Medlock2025}. As galaxies grow, supernovae and active galactic nuclei (AGN) are expected to inject significant amounts of energy and metals into the CGM via feedback, heating the gaseous halo, disrupting inflows, and changing the balance of phases \citep{SomervilleDave2015}. 

At the largest halo masses, multiphase structure is suppressed, and galactic atmospheres become dominated by a more uniform, hot, X-ray emitting intracluster medium (ICM), as observed in galaxy clusters at low redshift \citep{Voit2005,KravtsovBorgani2012}. In objects like these, cold gas survival is difficult outside of the constituent galaxies themselves, leading to a markedly different environment for galaxy evolution \citep{Lopez2008,Burchett2018,Ge2018,Lee2021}. Understanding when and how the multiphase CGM gives way to this hotter, enriched medium---the birth of the ICM---is a central problem in the formation of galaxy clusters.

The CGM to ICM transition is challenging to observe directly, but by combining both absorption spectroscopy and spatially resolved emission we can probe this evolution across gas phases. Absorption line spectroscopy of bright background sources is the workhorse of CGM studies at lower redshift, and has revealed large reservoirs of metal-enriched gas across ionisation states \citep[e.g.][]{Tumlinson2011,Werk2013,Prochaska2017,Lee2021,Anand2022}. The technique is sensitive to low column densities of intervening ions, but only provides data for a limited number of sight-lines, depending on the availability of background sources. In contrast, emission offers spatial information about the relatively diffuse gas surrounding galaxies, at the expense of detectability. Emission scales with the square of the density, making it particularly difficult to detect in the diffuse CGM. However, the advent of integral field units (IFUs) on large telescopes has revolutionised this field, with the detection of extended emission nebulae around galaxies now fairly commonplace \citep[e.g.][]{Cantalupo2014,Borisova2016,Wisotzki2016,Wisotzki2018,ArrigoniBattaia2019,Guo2020,MentuchCooper2026}. Some of the largest, so-called ``enormous'' Lyman-$\alpha$ (Ly$\alpha$) nebulae (ELANs) can be in excess of 100\,kpc in size, though these are rarer \citep{Cai2017,ArrigoniBattaia2019,ArrigoniBattaia2022,Li2024}. 

Ly$\alpha$ is the brightest line detected in emission in gaseous haloes, because of the abundance of hydrogen and the resonant nature of the line. These properties make Ly$\alpha$ easier to detect at larger radii from galaxies, but complicates the inference of the properties of the galactic atmosphere itself \citep{Dijkstra2014}. An alternative line to provide a cleaner probe of the CGM is H$\alpha$, as this traces only recombinations and is not affected by resonant scattering. Though H$\alpha$ is exceptionally difficult to detect from the ground at intermediate redshift due to atmospheric lines \citep{Leibler2018,Langen2023}, in the era of \textit{JWST} it is starting to be used to better understand underlying CGM properties \citep{Peng2025,Durovcikova2025}. 

A particularly interesting epoch to investigate the transition from CGM to ICM is near to cosmic noon at $z\sim3$. This is a critical window where massive haloes are growing rapidly and feedback is expected to have a strong impact on the galaxy and its surroundings. It is also the redshift where Ly$\alpha$ becomes detectable with MUSE on the VLT \citep{Borisova2016,Wisotzki2016}, allowing the combination of both absorption and emission across a significant sample of haloes. $z\approx2-3$ is also the peak of quasar activity in the Universe, making the identification of massive haloes much easier than it would otherwise be \citep{Richards2006}. From clustering studies, the kinematics of companions, and the kinematics of Ly$\alpha$ nebulae themselves, quasar hosts are estimated to have average halo masses of $\sim\!10^{12.5}$\,M$_\odot$ \citep{Shen2007,ArrigoniBattaia2019,Farina2019,Fossati2021,Eltvedt2024}, which makes them the likely progenitors of galaxy clusters by $z=0$ \citep[e.g.][]{Chiang2013}. Ongoing quasar activity also means ongoing energy injection into their host galaxies and CGM, allowing such haloes to be a powerful laboratory into the impact of feedback. The recent detection of significant galaxy overdensities at intermediate and high redshifts, especially with \textit{JWST}, also opens up the ability to target protoclusters without active AGN, allowing us to get a less biased picture of the evolution of the CGM \citep{Overzier2016,McConachie2022,Lamperti2024,Li2025,Witten2025a,Witten2025b}. At low redshift, halo masses of around $10^{13-14}\,\mathrm{M}_\odot$ are also found to be those in which baryon redistribution is most efficient, suggesting this is a key regime in the process of feedback-regulated halo growth \citep{Amodeo2021,Zhang2024,Hadzhiyska2025a,Hadzhiyska2025b,Pandey2025}. 

Over the past decade, advances in cosmological hydrodynamical modelling have enabled detailed simulations of galaxy cluster formation, exemplified by projects such as Magneticum \citep{Dolag2016}, \textsc{bahamas} \citep{McCarthy2017}, The 300 \citep{Cui2018}, \textsc{Fable} \citep{FABLE1}, MillenniumTNG \citep{Pakmor2023}, \textsc{flamingo} \citep{Schaye2023} and TNG-Cluster \citep{Nelson2024}. These simulations now routinely reproduce many observables like the mass-luminosity relation, but can have very different underlying physical models. Given the low number densities of massive haloes, cosmological simulations typically rely on two complementary approaches: 1) by simulating very large cosmological boxes (e.g.~Magneticum, MillenniumTNG, \textsc{flamingo}), which provide large, unbiased statistical samples but are less able to capture the detailed structure of the ICM, or 2) through running a large sample of cosmological zoom-in simulations, selected from a large dark matter-only box (e.g.~\textsc{bahamas}, \textsc{Fable} and TNG-Cluster), which achieve higher resolution at the expense of cosmological representativeness, and therefore cannot be used for analyses requiring matter sampling across a wide dynamic range. Moreover, these zoom-in simulations have mostly been analysed at either low redshift ($z\lesssim2$) where the most observational data (especially from X-rays) exists \citep[e.g.][]{Rasia2014,Barnes2017,Bennett2022,Rohr2024}, or at high redshift ($z\sim6$), where protoclusters are the most likely hosts of the first generation of quasars \citep[e.g.][]{Sijacki2009,Costa2014,Lupi2019,Ni2020,Bennett2024}. A small number of studies have begun to explore the intermediate redshift regime \citep[e.g.][]{Byrohl2021,Rohr2025}, though there remains a relative gap for highly resolved protocluster simulations at $z\sim3$---the critical window for understanding the CGM to ICM transition---which we aim to address in this paper. 

Modelling massive protoclusters, and especially comparing to measurements in the observational space, is not trivial. Numerical resolution is a key concern, with multiple works highlighting how the cold gas content of the CGM can significantly change when numerical resolution is increased \citep{Hummels2019,vandeVoort2019,Suresh2019,Peeples2019,Bennett2020}. In this paper we employ a scheme to enhance CGM resolution to try to address this, at the expense of only running a single halo to just below $z=3$. Only a handful of works have attempted to make mock Ly$\alpha$ emission maps from simulations due to the additional complexity of modelling radiative transfer, scattering and dust \citep[e.g.][]{Cantalupo2005,Laursen2009,Kollmeier2010,Byrohl2021,Kimock2021,Costa2022,deBeer2023,Obreja2024}, though in recent years several powerful tools have been released allowing robust modelling of some of these key processes \citep[e.g.][]{SmithA2015,RASCAS}. Among them is the Cosmic Lyman-$\alpha$ Transfer (\textsc{colt}) code \citep{SmithA2015}, which we use to post-process our cosmological simulations and generate mock emission maps we then compare to observations.

In this work we simulate a single protocluster, that goes on to become a $10^{15}\,\mathrm{M}_\odot$ galaxy cluster at $z=0$, with very high resolution in the CGM. We follow this object as it undergoes major mergers and feedback episodes until $z\sim2.7$, and study how the thermodynamics and metal content of the gaseous halo evolves as the protocluster moves into the cluster regime. Crucially, we also predict how this affects observables---looking at Mg\,\textsc{ii}, C\,\textsc{ii}, O\,\textsc{vi}, and O\,\textsc{vii} in absorption and Ly$\alpha$ and H$\alpha$ in emission.

This paper is structured as follows. We introduce the simulations used throughout this work in Section~\ref{Section:Methods:Simulations}. In Section~\ref{Section:Analysis} we then describe the analysis we perform, before showing how the global gas properties of the forming cluster evolve in Section~\ref{Section:Global}. Section~\ref{Section:Ions} shows how this evolution is reflected in mock ion absorption data, and in Section~\ref{Section:LyaHaloes} we present our mock Ly$\alpha$ haloes and compare our results to observational data. We discuss the implications and caveats of our results in Section~\ref{Section:Discussion}, before summarising our conclusions in Section~\ref{Section:Conclusions}.

\section{Simulations}
\label{Section:Methods:Simulations}

Our simulation uses the cosmological, hydrodynamical moving-mesh code \textsc{Arepo} \citep{Arepo}, which solves the Euler equations across a quasi-Lagrangian Voronoi mesh. We use cosmology consistent with Planck \citep{Planck2015Parameters}, where $\Omega_\mathrm{\Lambda} = 0.6911,$ $ \Omega_\mathrm{m} = 0.3089,$ $ \Omega_\mathrm{b}=0.0486,$ $\sigma_\mathrm{8} = 0.8159,$ $n_\mathrm{s} = 0.9667$ and $h=0.6774$. On-the-fly friends-of-friends \citep{FOF} and \textsc{Subfind} halo identification algorithms \citep{Subfind1,Subfind2} are used to identify gravitationally bound structures, and an on-the-fly shock finder is also used \citep{SchaalSpringel2015}. 

The simulation presented in this work is very similar to the ``ShockRef512" zoom simulation previously presented in \citet{Bennett2020}, but instead of stopping at $z=6$ we continue the simulation to approximately $z=2.7$. To begin with, we selected a massive galaxy cluster progenitor from the \textsc{Fable} simulation suite described in \cite{FABLE1}, itself selected from a parent dark matter only simulation, Millennium XXL \citep{MilleniumXXL}. We chose a halo with a mass at $z=0$ of $M_\mathrm{200} \approx 10^{15}$\,$\mathrm{M}_\mathrm{\odot}$\footnote{$M_\mathrm{200}$ and $M_\mathrm{500}$ refer to the mass contained within $R_\mathrm{200}$ and $R_\mathrm{500}$, respectively, within which the average density is $200$ and $500$ times the critical density of the Universe at the redshift of interest.}. This massive halo is ideal for the study of the evolution of the CGM into ICM at intermediate redshift. By the end of the simulation shown in this work, at $z=2.7$, the halo has a mass $M_\mathrm{200} \approx 7 \times 10^{13}$\,$\mathrm{M}_\mathrm{\odot}$, with a central stellar mass of $\sim\!5\times10^{11}\,\mathrm{M}_\odot$. The central galaxy is in the process of quenching, with recent feedback causing a rapid decrease in the star formation rate (SFR) to a specific SFR of $10^{-0.5}\,\mathrm{Gyr}^{-1}$.

The zoom-in region contains collisionless dark matter particles, extending to approximately $10$~cMpc, each with a mass of $m_\mathrm{DM} = 5.54 \times 10^7 \,h^{-1}\,\mathrm{M}_\mathrm{\odot}$. Baryonic cell masses are kept within a factor of two of a target mass, which is set depending on the shock refinement scheme described in \citet{Bennett2020}. Cells with a Mach number $\mathcal{M}>1.5$ are flagged for refinement, along with all cells within the shocked cell's smoothing length (except star-forming cells), for 10\,Myr. Shocks with $\mathcal{M}_\mathrm{thresh}>1.5$ are very prevalent throughout the forming protocluster, which effectively increases resolution throughout the zoom in region ($\gtrsim 95$ per cent of the volume within $R_{200}$ is at the highest resolution). To avoid refining in regions too far outside the halo of interest we also impose an additional radial cut on the resolution criterion, using the central black hole to track the halo centre. We set the cut off radius to $1.5$\,cMpc (approximately $R_{500}$ at $z=0$), larger than that used in \citet{Bennett2020}. The simulation is restarted from a pre-existing run with \textsc{Arepo}'s default refinement scheme at $z=12$, and the shock refinement scheme is activated at $z\sim10$.

The base baryonic resolution of the zoom in region has a target cell mass of $1.64 \times 10^{7}$\,M$_\odot$, with our boosted resolution a factor of 512 higher, $3.2 \times 10^4$\,M$_\odot$.  The increase in spatial resolution is a factor of $\sim\!8$, corresponding to the average cell radius in the CGM being $200-300$\,pc at $z\sim3$. This is therefore one of the highest resolution simulations of a halo of this size to date. 

We ran simulations with a slightly modified version of the \textsc{Fable} model, itself a modified version of the Illustris model \citep{Illustris1,Illustris2, Illustris3}. In brief, stellar feedback is implemented with hydrodynamically decoupled wind particles \citep[see][]{Illustris1}, with a third of the input wind energy imparted as thermal energy and the remainder is kinetic. AGN feedback has two modes---a ``quasar'' mode at high accretion rates, which injects thermal energy into the black hole's surroundings with an input duty cycle of 25~Myr, and a ``radio'' mode \citep[based on][]{Sijacki2007}, which injects bubbles into the inner CGM when the black hole mass increases by 1 per cent. For full details of the \textsc{Fable} model see section~$2$ of \citet{FABLE1}. 

Like in \citet{Bennett2020}, we only include a single, central black hole in our halo of interest, which is seeded slightly later than its equivalent in \textsc{Fable}. This black hole acts as a tracer of the halo's potential minimum when the new shock refinement scheme is activated \citep[as used in][]{Suresh2019}. As a consequence of this, the black hole does not have a contribution to growth from mergers, meaning the black hole masses in our shock refined runs lie approximately 20 per cent below those in \textsc{Fable} by $z=2.7$, with an equivalent decrease in the injected feedback energy. Despite this difference, the central stellar mass of the protocluster at $z=2.7$ differs by less than 5 per cent between the two runs. As we wish to compare the nebular emission properties of the haloes on a similar basis, for our radiative transfer calculations (see Section~\ref{Section:Methods:Colt}) we assume both haloes to have the same black hole mass as our high resolution simulation for the purposes of calculating the Eddington luminosity. In this work we only use the fiducial \textsc{Fable} run to investigate the impact of resolution in Section~\ref{Section:Resolution}, all other results use our high resolution CGM simulation. 

A second difference we have from \textsc{Fable} is that the UV background of \citet{FG_UVB} is active from around $z\sim10.6$, rather than activated instantaneously at $z=6$ like in \textsc{Fable} and Illustris. We note that the true reionisation history of such a halo is likely not captured by a spatially uniform UV background, as reionisation topology should proceed very differently in an overdense environment and would likely occur at $z>10$. Finally, to somewhat compensate for the changes in resolution we increase the number of neighbouring cells that feedback energy and metals are injected by a factor of $128$, as in \citet{Bennett2020}. All other parameters are kept the same as in \textsc{Fable}.

\section{Analysis} \label{Section:Analysis}

In this work we explore mock observations in both absorption and emission in the gaseous halo of the growing protocluster. For most of the paper we focus on gas ionisation states set via photoionisation equilibrium with the metagalactic UVB or with local stellar sources, which we find to be similar. The similarity between these ionisation prescriptions allows us to isolate the hydrodynamical effects of cluster growth on the cold gas content. Later, we discuss the impact of AGN ionisation, which we will explore in more detail in future work. 

\subsection{Ion absorption with \textsc{trident}} \label{Section:Methods:Trident}

The most common way of studying the gas around galaxies is through absorption spectroscopy, where photons emitted from a bright background source may get absorbed by intervening material. In the first part of our analysis, we look at several ions commonly used in such studies---Mg\,\textsc{ii}, C\,\textsc{ii}, O\,\textsc{vi}, and O\,\textsc{vii}. The ionisation states of these elements are not natively tracked in the simulation, so we calculate them in post-processing using \textsc{trident} \citep{Trident}. It assumes ionisation equilibrium with the metagalactic UVB of \citet{FG_UVB}, taking into account self-shielding using the prescription of \citet{Rahmati2013}, and uses lookup tables of data from \textsc{cloudy} \citep{Cloudy}. While this likely underestimates the absolute level of metal ionisation in our object due to a lack of local radiation sources (as discussed later), the relative changes in ion prevalences can still help us interpret observations of the cluster's cold gas content. 

\subsection{Nebular emission with \textsc{colt}} \label{Section:Methods:Colt}

To explore the properties and evolution of the nebular emission surrounding our halo, we utilise the Cosmic Lyman-$\alpha$ Transfer (\textsc{colt}) code \citep{SmithA2015}. \textsc{colt} is a Monte Carlo radiative transfer (MCRT) solver, which we apply in post-processing to our simulation output to calculate photoionization and line emission. We run \textsc{colt} within a spherical region extracted from a fixed physical radius of 500\,kpc in each snapshot, centred on the largest subhalo (and the central black hole) at that time.

Photon packets are spatially sampled based on the luminosity of our ionising sources and launched isotropically from each source. These then propagate through the recreated Voronoi mesh of the gas cells in the simulation. As they propagate, they interact with the surrounding gas cells through photoionisation, as well as scattering and attenuation by dust. In \textsc{Fable} we do not have on-the-fly dust modelling, so we explore several input dust-to-metal (DTM) ratios following a Milky Way dust law \citep{WeingartnerDraine2001}. We also tested a SMC dust law, but found that the results showed little difference with respect to the Milky Way dust law assumption. By $z=2.7$, the metallicity of the central galaxy’s ISM is $0.45\,Z_\odot$. A Milky-Way-like dust-to-metal ratio of $\mathrm{DTM}\sim0.3$–0.4 is therefore a plausible comparison. However, the effective dust abundance relevant for Ly$\alpha$ radiative transfer is highly uncertain in our model because the multiphase ISM is unresolved and dust growth, destruction, and expulsion by stellar and AGN feedback are not followed self-consistently. We therefore explore a range of DTM ratios (0, 0.1, and 0.3) as a systematic uncertainty rather than as predictions of the simulation. Higher DTM values further suppress the central Ly$\alpha$ emission, while lower values provide a better match to observed central surface brightnesses, although this remains degenerate with unresolved ISM structure and AGN-powered Ly$\alpha$ production.

Because the opacities of simulated gas cells are dependent on the gaseous ionisation state, \textsc{colt} iterates between the MCRT calculation (with $10^8$ photon packets) and abundance updates until the global H\textsc{ii} recombination rate is converged to within 0.1 per cent. To avoid unphysical artefacts due to the abrupt edge of our extracted sphere, we only allow emission of photons from sources within a radius of 400\,kpc, with photons allowed to freely escape once they reach a radius of 450\,kpc.

For this study we produce maps of H$\alpha$ and Ly$\alpha$ emission around our simulated protocluster, and as such only follow the ionisation of H and He; in future works we will expand this to study both metal emission and a detailed investigation of the kinematics of these nebulae. Below, we briefly summarise the setup we adopt for \textsc{colt}. For more details on the code see \citet{SmithA2015,SmithA2017,SmithA2019,SmithA2022,SmithA2025}, and for details of the most recent version used in this work see \citet{McClymont2025}.

\begin{figure*}
    \centering
    \includegraphics[width=0.9\linewidth]{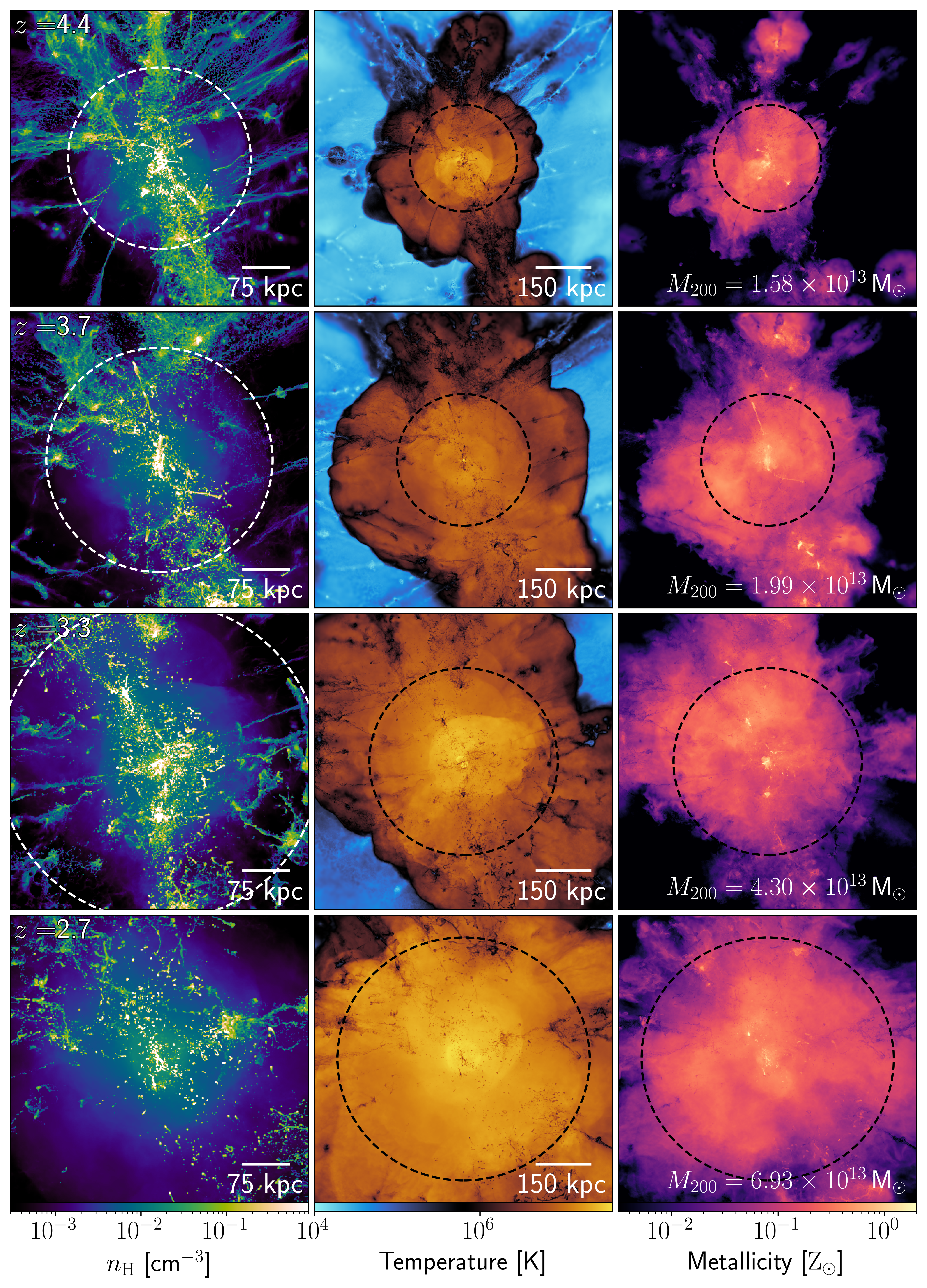}
    \caption{Maps of gas number density (left column), temperature (centre column) and metallicity (right column), at four redshifts $z = 4.4, 3.7, 3.3,$ and $2.7$ (from top to bottom). All maps show mass-weighted average quantities. The density maps are zoomed in further to highlight the transition from a filamentary to clumpy cool gas structure in the CGM as the halo grows. Dashed circles in all panels denote $R_{200}$.}
    \label{fig:maps}
\end{figure*}

\subsubsection{Ionising sources}

In all analysis with \textsc{colt} in this work we include stars as sources and use spectral energy distributions (SEDs) from BPASS \citep[v2.2.1,][]{BPASS2017}. To better sample emission from lower luminosity sources, we use a luminosity boosting that biases the MCRT source probability distribution and weights by a power law proportional to $L^{\beta}$, with $\beta=0.75$. We also enforce a minimum energy bin width of 0.05~dex for the radiation transport. Like the native \textsc{Fable} simulation, our \textsc{colt} analysis also includes a uniform UV background including effects from self-shielding, following \citet{FG_UVB} and \citet{Rahmati2013}. We also include line emission due to collisional excitation (for gas that is not star forming, see below), using analytic expressions from \citet{CenOstriker1992} and \citet{SmithA2022}. 

In our simulation we do not resolve the multiphase structure of the ISM, and rely on an effective equation of state model based upon \citet{SpringelHernquist2003}, in which the temperature of star-forming gas is not self-consistently tracked. Using the output temperatures of these cells for ionisation calculations can therefore lead to very unphysical results. To alleviate this, we enforce a constant temperature of $10^4$\,K for star-forming gas, and do not include line emission due to collisional excitation for this gas \citep[photoionisation from young stars is dominant and properly accounted for anyway, see e.g.][]{Byrohl2021}. This may lead to an underestimate in the Ly$\alpha$ luminosity of ISM gas, which is degenerate with our dust modelling, although again relative changes down to the ISM density threshold can still provide insights into the changing conditions inside growing protoclusters.

In the later parts of our analysis we also include a contribution from AGN. For this work we only consider radiation coming from a single central black hole in the halo that is injected isotropically. Previous versions of \textsc{colt} used three fixed radiation bins for the AGN contribution, but here we extended the implementation to use an arbitrary SED over the same spectral grid as the stellar contribution---for this work we only consider the ionisation of hydrogen and helium, and will explore metal ionisation in future work. To do this, we take the SED model from \citet{Shen2020}, linearly interpolate their values for $\log(\nu L_\nu [\mathrm{erg}\,\mathrm{s}^{-1}])$ onto a logarithmic wavelength grid, and convert to $L_\lambda$. We then normalise the spectrum such that $\int_0^\infty L_\lambda d\lambda = 1$. This SED is then scaled with the input bolometric luminosity. In this work, when an AGN contribution is included this input luminosity is taken to be the Eddington luminosity of the central black hole in the high resolution simulation (see Section~\ref{Section:Methods:Simulations}). Combined with our results without any AGN ionisation, these two models bracket the likely range of ionisation fields a protocluster would have.

\begin{figure}
    \centering
    \includegraphics[width=\linewidth]{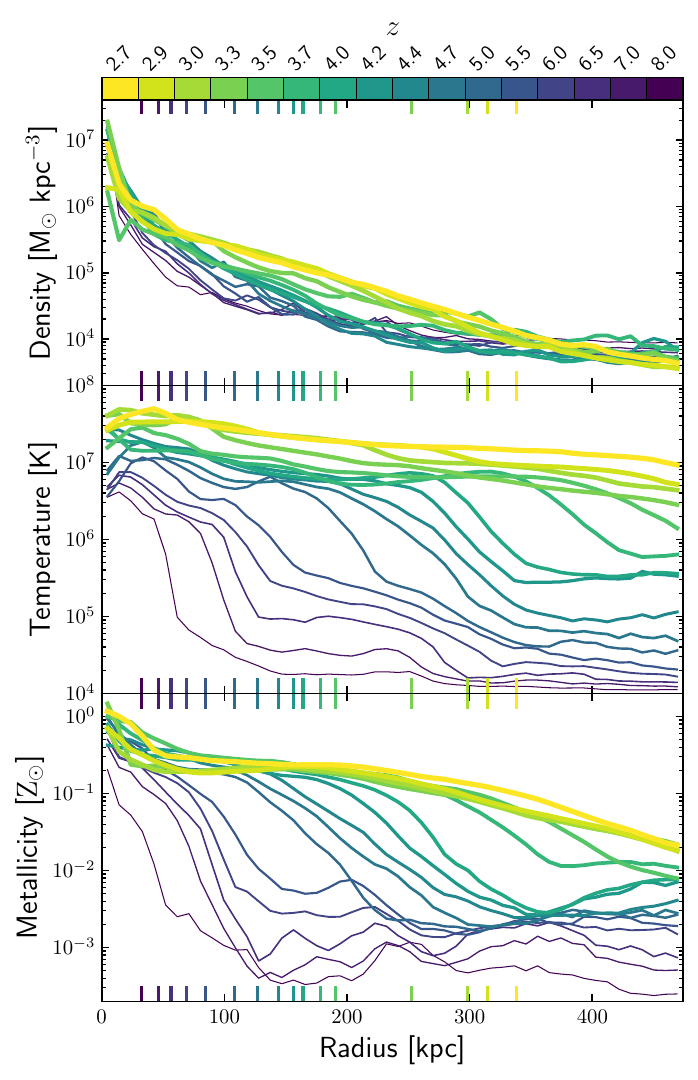}
    \caption{Volume-weighted average profiles of gas density (top panel), temperature (middle panel) and metallicity (bottom panel) for the protocluster from $z=8$ (purple) to $z=2.7$ (yellow). Coloured markers at the top and bottom of each panel show the position of $R_{200}$ at each time. As the halo grows we can see how the density profile flattens, the hot halo expands, and the forming ICM is enriched to a near-uniform metallicity.}
    \label{fig:profs}
\end{figure}

\begin{figure}
    \centering
    \includegraphics[width=\linewidth]{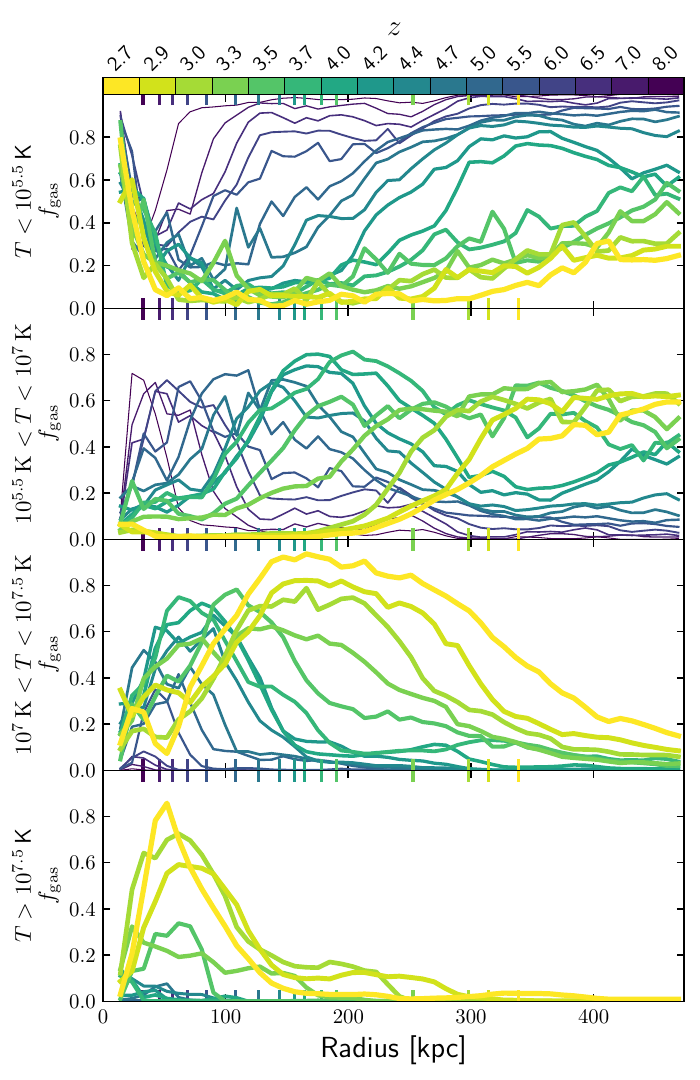}
    \caption{Radial profiles of gas fractions in different phases for the protocluster from $z=8$ (purple) to $z=2.7$ (yellow). The panels show the fractions for, from top to bottom, gas with temperature $T<10^{5.5}\,\mathrm{K}$ (including all star-forming gas), $10^{5.5}\,\mathrm{K}<T<10^7\,\mathrm{K}$, $10^{7}\,\mathrm{K}<T<10^{7.5}\,\mathrm{K}$, and $T>10^{7.5}\,\mathrm{K}$. Coloured markers at the top and bottom of each panel show the position of $R_{200}$ at each time. The gaseous halo heats up from the inside out as it grows, with the increasing temperature affecting the ionisation state of the nascent ICM.}
    \label{fig:gasphasefracs}
\end{figure}

\section{Global evolution of thermodynamic protocluster properties}
\label{Section:Global}

We start by exploring the evolution of the thermodynamic properties of the protocluster as it grows, from $z=8$ to the end of the simulation at $z=2.7$. We note that we show all maps and radial profiles in physical coordinates at all times, to highlight how the CGM grows and develops into the ICM.

\subsection{Maps of key protocluster thermodynamic properties}

Firstly, in Fig.~\ref{fig:maps}, we show maps of mass-weighted average density, temperature and metallicity at four redshifts spanning the development of the ICM. Every temperature and metallicity panel has a fixed width of 800\,pkpc, with the density panels zoomed in slightly to highlight the cold gas structure in the centre. 

In the top row at $z=4.4$, we see the protocluster already has a sizeable metal-enriched hot halo in place extending beyond $R_\mathrm{200}$, though numerous cooler filaments feed material towards the centre of the halo. Cool, dense gas is abundant in the multiphase CGM, despite ongoing quasar-mode AGN feedback (visible as the higher central temperatures). There is plenty of primordial gas present in the outskirts, though metals from both the protocluster and infalling galaxies are actively polluting the intergalactic medium (IGM).

By $z=3.7$, $\sim\!315$\,Myr later, we already see a significant expansion of the hot halo of the protocluster. This engulfs many inflowing filaments, seen particularly clearly in the temperature map. While these filaments remain cool whilst embedded in the hot halo, they appear less able to penetrate within $R_\mathrm{200}$, with most becoming disrupted before reaching the virial radius. However, as seen in the left-hand panel, significant amounts of dense gas are still present in the CGM. In the bottom of this panel, we can also see another massive halo beginning to merge, which will strongly affect the forming ICM.

The third row, a further $\sim\!260$\,Myr later at $z=3.3$, captures a moment in the middle of this major merger, which has a mass ratio of approximately 1:2.5. The temperature of the volume filling phase has increased near the cluster centre, due to the dual effect of a merger shock (visible as a discontinuity in temperature) and radio-mode AGN feedback (visible in the very centre of the halo). Despite this widespread hot gas component, a huge amount of cool, dense gas remains throughout the protocluster. We can see how cool gas has a broad range of densities at this point, which we will explore in more detail using specific ion tracers in Section~\ref{Section:Ions}. The metallicity map is already fairly uniform by this time, though there are still pockets of low metallicity gas within the central $\sim\!200$\,kpc.

By the end of the simulation at $z=2.7$, an additional $\sim\!440$\,Myr later, the properties of the forming cluster have changed significantly. The bulk of the gas has been heated above $\sim\!3\times10^7$\,K, with additional energy still being injected by the central black hole. While some cool gas still exists embedded in the hot halo, it is now primarily constrained to the highest density clumps---the constituent galaxies of the cluster themselves. As we discuss in Section~\ref{Section:Ions}, the covering fraction of intermediate density gas drops sharply by $z=2.7$. The metallicity map is also much smoother across a large radial range, and particularly around the central galaxy, widely reaching the canonical $\sim\!0.3\,\mathrm{Z}_\odot$ observed in lower redshift galaxy clusters.

In this massive halo at $z\sim3$, the combined impact of a major merger and powerful AGN feedback drives a transformation of the CGM---from the inside out---into a hot, volume-filling atmosphere resembling the ICM seen at low redshift. Determining whether such inside-out CGM heating is ubiquitous among $\sim\!10^{13}\,M_\odot$ haloes at $z\sim3$, or is instead contingent on the specific merger and AGN history of this system, will require a broader suite of high-resolution protocluster simulations.

\subsection{Thermodynamic radial profiles of the gaseous halo}

We now turn to how the growth of the protocluster affects the radial distribution of gas properties. In the top panel of Fig.~\ref{fig:profs}, we show the volume-weighted average gas density profile of the developing protocluster, over a longer timeframe than the maps shown previously. The volume weighting highlights the properties of the hot, volume-filling gas phase, as well as somewhat downplaying changes due to merging substructure. The shape remains similar for much of the protocluster's growth, with a sharp drop-off within 100\,kpc and a gradual flattening. A noticeable change comes in the $50-250\,$kpc radial range towards the end of the simulation: material is redistributed outwards from the inner parts of the cluster. This is most likely due to a combination of the merger activity and the radio-mode AGN feedback in \textsc{Fable}, which injects $\sim\!3\times10^{60}$\,erg of energy into the forming ICM in less than 500\,Myr before the end of the simulation. Previous work with \textsc{Fable} has shown the model can redistribute material to larger radii, albeit at lower redshift \citep{Bennett2022}.  

The injection of energy into the halo is also clearly visible in the volume-weighted average temperature profiles in the middle panel of Fig.~\ref{fig:profs}. These display a much more significant evolution with redshift, showing the growth of the hot halo over time as the virial shock expands outwards (also visible in Fig.~\ref{fig:maps}). Inside of the virial shock we also see a gradual heating of gas as the virial temperature also increases \citep[though we note the virial temperature is often not a good description of the whole halo due to non-thermal motions, particularly when low-density gas is better resolved, see e.g.][]{Vazza2018,Lochhaas2021,Bennett2022}. The radius of the temperature increase from the IGM to the CGM also grows as the virial shock around the halo expands. This accretion shock typically lies between 1.5 and $2\,R_{200}$, and is also clearly visible in the central panels of Fig.~\ref{fig:maps}. Focussing again on the $50-250\,$kpc radial range, we can see how the bulk of gas is heated from $\sim\!10^7$\,K to above $3\times10^7$\,K around $z\sim3$, at the same time as the average density of the gas is increasing. We explore how this affects the phase structure of the gas below, and how that influences particular ions in the following Section. 

As discussed earlier, the redistribution of mass outwards due to feedback also disperses metals throughout the forming cluster, and the maps in Fig.~\ref{fig:maps} show how the forming ICM mixes and becomes more uniform during this process. In the bottom panel of Fig.~\ref{fig:profs} we show the radially averaged, volume-weighted metallicity profiles. We see how metal-poor gas dominates the CGM and IGM at high redshifts, which is gradually enriched via stellar winds and supernovae, and transported further outwards with the aid of quasar feedback. As the protocluster develops, the metallicity profile gradually flattens from the inside out to the canonical value of $\sim\!0.3\,\mathrm{Z}_\odot$. We note that significant enrichment at a larger distance from the central galaxy ($\sim\!350$\,kpc) only occurs after the onset of radio-mode feedback in \textsc{Fable} at $z\lesssim3.5$---the flatness of the metallicity profile in the outskirts of forming galaxy clusters is a good indicator of the presence of powerful, ejective feedback.

In Fig.~\ref{fig:gasphasefracs} we present radial profiles of gas mass fractions split by temperature. The top panel shows gas with $T < 10^{5.5}$\,K, which includes star-forming gas. The ISM is clearly present at small radii throughout the simulation, however at high redshift we can also see this cool gas dominating the IGM. As the halo grows, the cool gas mass fraction at fixed radius drops. This proceeds outwards---the inner CGM is depleted of cool gas first, but eventually cool gas only contributes a quarter of the total CGM mass even at $500$\,kpc. 

The second panel includes gas with intermediate temperatures between $10^{5.5}$\,K and $10^{7}$\,K. At every redshift, the profile of this gas increases from near zero in the ISM to $>50$ per cent in the CGM, before dropping off again. However, the radius at which the fraction peaks shifts outwards dramatically with time. At a radius of $\sim\!180$\,kpc, this warm gas makes up 80 per cent of the mass of the CGM at $z\approx4$. By $z=2.7$, this fraction is nearly zero. Conversely, at a radius of $\sim\!350$\,kpc the fraction of gas in this temperature range increases, as the merger and feedback heat cool gas in the outer CGM. 

Within $\sim\!200$\,kpc, by $z\sim3$ there is very little mass left in the CGM with a temperature less than $10^7$\,K. The third row shows gas between $10^7$\,K and $10^{7.5}$\,K, and we can see this gas makes up the largest share of the mass fraction at around $100$\,kpc at $z\sim3.5$. However, after the major merger and subsequent feedback, the inner halo is also depleted of gas at this temperature, while it dominates the hot halo between $150$\,kpc and $250$\,kpc by $z=2.7$. 

The inner part of the gaseous halo is heated even further, beyond $10^{7.5}$\,K, by the end of the simulation, as seen in the bottom panel of Fig.~\ref{fig:gasphasefracs}. This extremely hot gas is only present in significant quantities after $z\sim3$, and forms the strong X-ray emitting bulk of the newly formed ICM. Such dramatic shifts in the phase structure of the halo over a relatively short period of time will have strong effects on observational probes, which we now explore. 

\section{Evolution of ions} \label{Section:Ions}

The changes in the thermodynamic properties of the halo described in Section~\ref{Section:Global} lead to changes in a wide number of gas observables. We looked at a number of different ion tracers, many of which shared similar evolutions. In this paper we therefore focus on a handful of ions commonly used to investigate different phases of circumgalactic gas---Mg\,\textsc{ii} and C\,\textsc{ii} (from gas with density $n_\mathrm{H}\gtrsim10^{-2}\,\mathrm{cm}^{-3}$ and temperature around $T\approx10^{4-4.5}$\,K), O\,\textsc{vi} (from gas with typical density $n_\mathrm{H}\approx10^{-5}\,\mathrm{cm}^{-3}$ and temperature $T\approx10^{5.5}$\,K), and O\,\textsc{vii} (from gas with density $-6\lesssim \log(n_\mathrm{H}/\mathrm{cm}^{-3})\lesssim -4$ and temperature peaking at $T\approx10^6$\,K). 

\subsection{$\mathrm{Mg}$\,\textsc{ii} \& $\mathrm{C}$\,\textsc{ii}---the depletion of cold gas in the forming ICM}

\begin{figure}
    \centering
    \includegraphics[width=\linewidth]{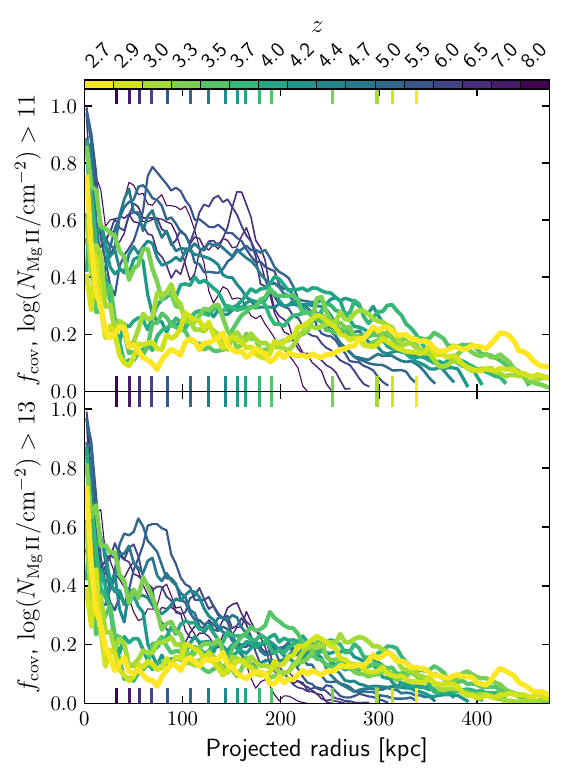}
    \caption{Evolution of the covering fraction of Mg\,\textsc{ii}, for column densities $N_{\mathrm{Mg\,II}} > 10^{11}$\,cm$^{-2}$ (top panel), and for columns $N_{\mathrm{Mg\,II}} > 10^{13}$\,cm$^{-2}$ (bottom panel) as a function of radius. Coloured markers at the top and bottom of each panel show the position of $R_{200}$ at each time. The covering fraction of Mg\,\textsc{ii} drops with time as the halo grows, with the $50-250\,$kpc range particularly affected for high column densities as intermediate density gas is destroyed.}
    \label{fig:MgIICoverFrac1e11}
\end{figure}
\begin{figure}
    \centering
    \includegraphics[width=\linewidth]{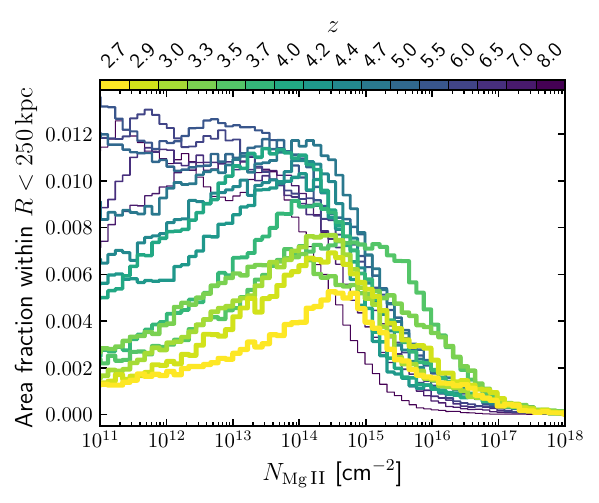}
    \caption{Evolution of the distribution of Mg\,\textsc{ii} column densities within 250\,kpc of the forming cluster's centre. Intermediate columns of Mg\,\textsc{ii} are gradually destroyed, while the highest columns remain largely unchanged, leading to a clumpier distribution dominated by satellite galaxies.}
    \label{fig:MgIIHist}
\end{figure}
\begin{figure}
    \centering
    \includegraphics[width=\linewidth]{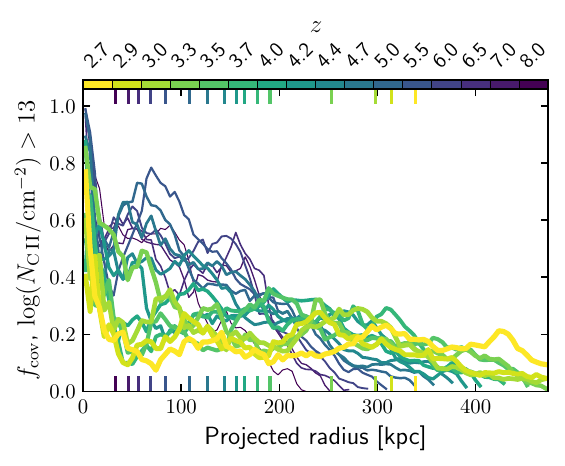}
    \caption{Evolution of the covering fraction of C\,\textsc{ii}, for column densities $N_{\mathrm{C\,II}} > 10^{13}$\,cm$^{-2}$ as a function of radius. Coloured markers at the top and bottom of each panel show the position of $R_{200}$ at each time. C\,\textsc{ii} shows a very similar evolution to Mg\,\textsc{ii}, albeit at slightly higher column densities.}
    \label{fig:CIICoverFrac1e13}
\end{figure}

In the top panel of Fig.~\ref{fig:MgIICoverFrac1e11} we show the evolution of the covering fraction of Mg\,\textsc{ii} with column density $N_{\mathrm{Mg\,II}} > 10^{11}$\,cm$^{-2}$ within 250\,kpc around the forming protocluster centre. We can see the gradual decline of Mg\,\textsc{ii} in the CGM, particularly in the radial range $50-250$\,kpc, with the covering fraction dropping from $\sim\!70$ percent at $z>6$ to $\sim\!15$ percent at $z\sim3$. Much of the early evolution is likely driven by both the increasing virial temperature of the halo and quasar-mode feedback. Interestingly, we see a temporary enhancement of the Mg\,\textsc{ii} covering fraction within $100$\,kpc at $z=3.3$, which is most likely due to tidal debris from the merger shown in the third row of Fig.~\ref{fig:maps}. After this, higher ambient temperatures in the gaseous halo, due to the merger and radio-mode AGN feedback, destroy much of the cold gas in the halo. The bottom panel shows a higher column density threshold, $N_{\mathrm{Mg\,II}} > 10^{13}$\,cm$^{-2}$, which leads to a similar evolution but with the change restricted to smaller radii.

By the end of the simulation at $z=2.7$ the covering fraction is predominantly due to the surviving dense clumps, often associated with the constituent galaxies of the protocluster. We see this in Fig.~\ref{fig:MgIIHist}, which shows the evolution of the area covered by a given column density within 250\,kpc of the halo centre, normalised by the total area within that radius. At the densest end, above $N_{\mathrm{Mg\,II}} \gtrsim 10^{15}$\,cm$^{-2}$, the coverage remains largely unchanged with time, as the feedback is not powerful enough to disrupt the densest clumps. The only notable change at these very high columns is during the merger, where dense Mg\,\textsc{ii} is significantly enhanced. At lower columns, in the range $10^{11}\lesssim N_{\mathrm{Mg\,II}} \lesssim 10^{13}$\,cm$^{-2}$ the fraction of sight-lines particularly drops after the merger and activation of radio-mode feedback at $z \lesssim 3$. 

In Fig.~\ref{fig:CIICoverFrac1e13} we additionally show the evolution in covering fraction of C\,\textsc{ii} above $N_{\mathrm{C\,II}} \gtrsim 10^{13}$\,cm$^{-2}$. This shows a very similar evolution to Mg\,\textsc{ii}, albeit at a slightly higher column density.

Throughout the evolution of this halo, we see a general trend of intermediate columns becoming less frequent, with Mg\,\textsc{ii} and C\,\textsc{ii} further ionised and distributed throughout the hot halo. What cool gas remains becomes increasingly clumpier as the protocluster develops into a young galaxy cluster. 

\subsection{$\mathrm{O}$\,\textsc{vi} \& $\mathrm{O}$\,\textsc{vii}---the increasing dominance of hot gas}

\begin{figure}
    \centering
    \includegraphics[width=\linewidth]{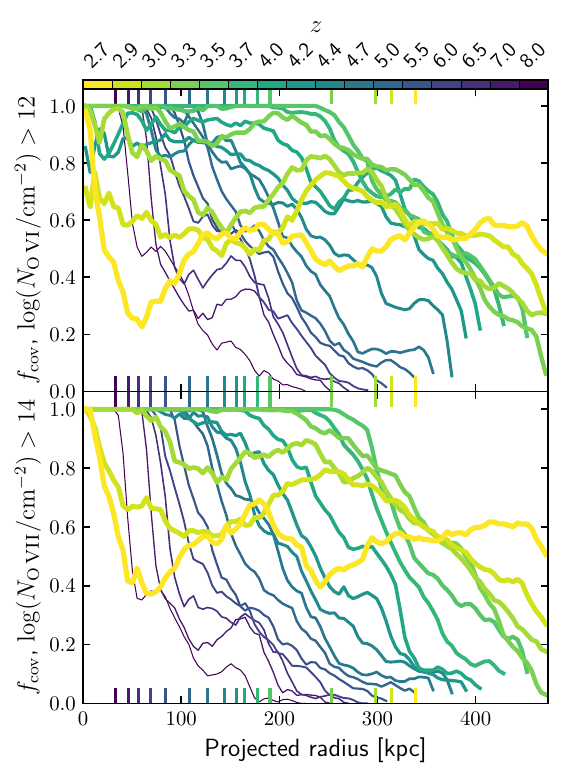}
    \caption{Evolution of the covering fraction of O\,\textsc{vi}, for columns $N_{\mathrm{O\,VI}} > 10^{12}$\,cm$^{-2}$ (top panel), and O\,\textsc{vii}, for columns $N_{\mathrm{O\,VII}} > 10^{14}$\,cm$^{-2}$ (bottom panel), as a function of radius. Coloured markers at the top and bottom of each panel show the position of $R_{200}$ at each time. Initially, the expansion of the hot halo allows oxygen to become ionised at larger radii, but by the end the gas is ionised to even higher states from the inside out.}
    \label{fig:OVIICoverFrac1e14}
\end{figure}

\begin{figure}
    \centering
    \includegraphics[width=\linewidth]{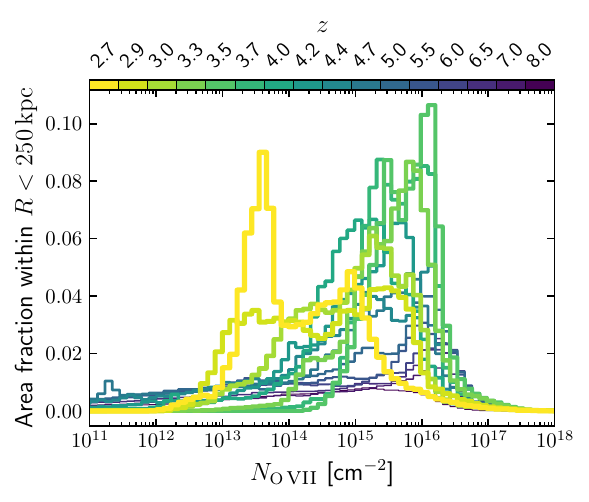}
    \caption{Evolution of the distribution of O\,\textsc{vii} column densities within 250\,kpc of the forming protocluster. High column densities build up as the halo grows and ionises more O\,\textsc{vii}. However, a bimodal distribution by $z=2.7$ shows how gas begins to become ionised beyond O\,\textsc{vii} as the ICM forms.}
    \label{fig:OVIIHist}
\end{figure}

We have shown how the halo is depleted of Mg\,\textsc{ii} and C\,\textsc{ii} over time as cool, dense gas is heated and dispersed. This gas ends up in a higher ionisation state---here we focus on O\,\textsc{vi}, present at temperatures $\sim\!10^{5.5}\,$K, and O\,\textsc{vii}, with a temperature of $\sim\!10^{6}$\,K. In Fig.~\ref{fig:OVIICoverFrac1e14} we show the evolution of the covering fraction of O\,\textsc{vi} with columns $N_{\mathrm{O\,VI}} > 10^{12}$\,cm$^{-2}$ (top panel) and O\,\textsc{vii} with columns $N_{\mathrm{O\,VII}} > 10^{14}$\,cm$^{-2}$ (bottom panel). We see similar evolutions for both ions---at high redshift, such hot gas is localised within the immediate vicinity of the galaxy, heated by galactic winds and quasar feedback. Over time, the O\,\textsc{vi} and O\,\textsc{vii} grow as the hot halo grows, reaching $\sim\!250$\,kpc in size by $z=3.5$. Throughout this time, the covering fraction sharply drops off from unity, indicating the edge of the hot gaseous halo, with ionised oxygen ubiquitous within it. 

After $z=3.5$ we see a change in behaviour. The covering fraction---particularly in the $50-250$\,kpc range---drops. This occurs despite the average volume-weighted density in this region increasing, due to the increase in temperature driven by both the merger and radio-mode feedback (see Fig.~\ref{fig:profs}). Significant amounts of oxygen get further ionised beyond O\,\textsc{vii} as the forming protocluster's emission enters the strong X-ray regime. We show the evolution of the column density distribution in Fig.~\ref{fig:OVIIHist}, demonstrating that while O\,\textsc{vii} is still widely present in the halo, it becomes shifted to lower column densities as the dense cluster core is heated. 

\section{Ly$\alpha$ haloes} \label{Section:LyaHaloes}

\begin{figure}
    \centering
    \includegraphics[width=\linewidth]{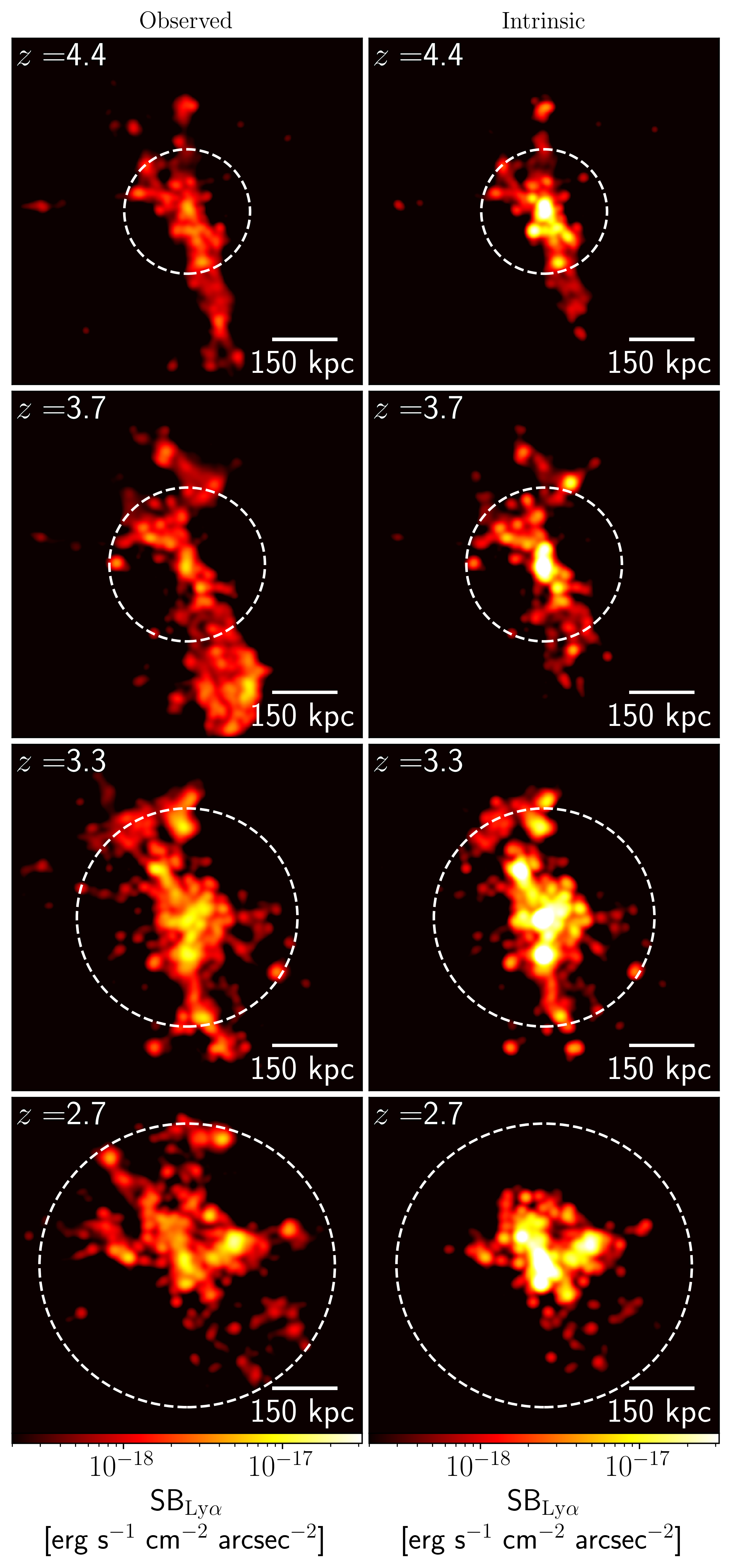}
    \caption{Spatial maps of Ly$\alpha$ surface brightness (with no AGN ionisation), convolved with a 1 arcsecond Gaussian, at four redshifts $z=4.4,3.7,3.3,$ and $2.7$ (from top to bottom). The left column shows observed maps, including the effects of dust attenuation and Ly$\alpha$ resonant scattering, and the right column shows intrinsic maps, showing pure emission without attenuation or scattering. The size and orientation of the projections are the same as shown in Fig.~\ref{fig:maps}. Dashed white circles in each panel show $R_{200}$. Ly$\alpha$ traces the presence of dense, neutral gas, and resonant scattering is required to illuminate halo outskirts from a central source.}
    \label{fig:Lyamaps}
\end{figure}

\begin{figure*}
    \centering
    \includegraphics[width=0.49\linewidth]{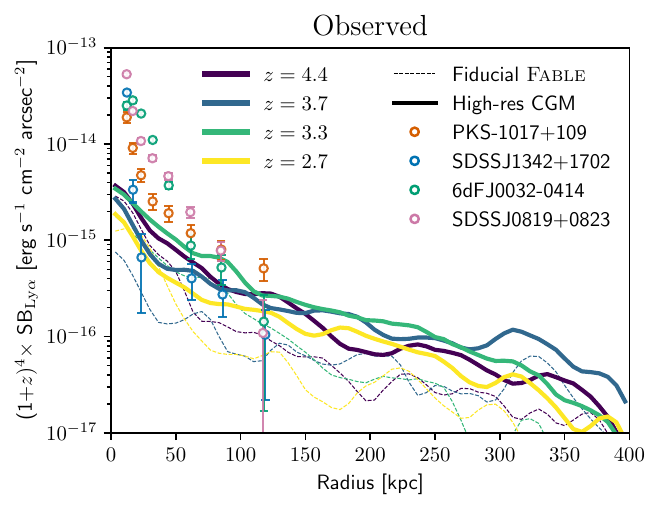}
    \includegraphics[width=0.49\linewidth]{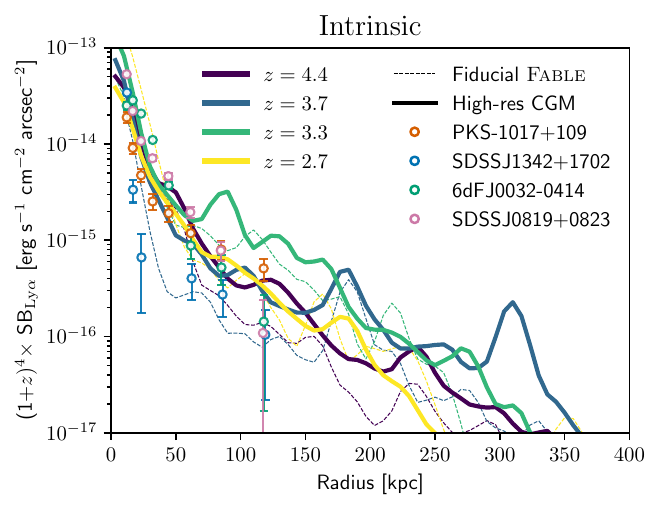}
    \caption{Radial profiles of Ly$\alpha$ surface brightness, corrected for the effect of cosmological dimming. Lines show median profiles of maps projected along the three axes of the simulation box---thin, dashed lines show the fiducial \textsc{Fable} simulation and thicker, solid lines for the high-resolution shock refined simulation. Coloured points show observational data for a few of the largest haloes in the QSOMuseum sample \citep{ArrigoniBattaia2019,GonzalezLobos2025}. The left panel shows the full mock profiles from \textsc{colt} with an assumed dust-to-metal ratio of $0.1$, while the right panel shows the intrinsic emission without dust attenuation or scattering.}
    \label{fig:LyaProfilesNoAGN}
\end{figure*}

While absorption spectroscopy is a sensitive diagnostic of gas thermodynamical state, it can only be used along (a few) lines-of-sight through a given halo. Alternatively, detecting emission is a promising way to explore the gas content of the CGM and emerging ICM---particularly using bright lines like Ly$\alpha$ and especially in massive haloes like the one we have simulated. In this section we generally refer to two outputs---`observed' and `intrinsic'. For the \textit{observed} maps and profiles, we include the full \textsc{colt} treatment of dust attenuation (with an assumed dust-to-metal ratio of 0.1), scattering off of dust, and Ly$\alpha$ resonant scattering. In contrast, the \textit{intrinsic} data shows only the raw emission from the simulation, without any attenuation or scattering. In Section~\ref{Section:Variations} we discuss further model variations, including a version with resonant scattering but no dust, and with the inclusion of AGN ionisation. 

\subsection{Ly$\alpha$ emission maps}

Using our output from \textsc{colt}, in Fig.~\ref{fig:Lyamaps} we show maps of the surface brightness (SB) of the halo's Ly$\alpha$ nebular emission, convolved with a 1 arcsecond Gaussian filter---approximately the average seeing of observations with MUSE \citep[e.g. QSOMuseum,][]{ArrigoniBattaia2019}. These maps are shown at the same redshifts and orientations as Fig.~\ref{fig:maps} for easy comparison. The bottom of the colour map is set to be a factor of a few below the typical sensitivity limit for observations---for example, the QSOMuseum sample has a $2\sigma$ limit of $8.8\times10^{-19}$\,erg\,s$^{-1}$\,cm$^{-2}$\,arcsec$^{-2}$ \citep{ArrigoniBattaia2019,GonzalezLobos2025}. We note Fig.~\ref{fig:Lyamaps} does not normalise for surface brightness dimming with redshift, and as such directly predicts the expected Ly$\alpha$ halo brightness at each snapshot. The left-hand column shows the observed outputs from the simulation, and the right-hand column shows the intrinsic Ly$\alpha$ emission (both described at the beginning of Section~\ref{Section:LyaHaloes}). We first present the observed output below.

In the top left panel of Fig.~\ref{fig:Lyamaps} at $z=4.4$ we can clearly see the large scale structure of the cosmic web, with elongated emission tracing the main filament hosting the halo. Some additional substructure is also present in the lower luminosity parts of the halo. A bifurcation near $R_{200}$, where multiple filaments extend from the node of the halo, is also seen in the density plot in Fig.~\ref{fig:maps}. There are also signs of a horizontal filament to the left of the main halo, however only the densest parts of gas show up at this sensitivity. Identifying such a structure would therefore likely require longer exposure times than have typically been used so far. 

$\sim\!315$\,Myr later, at $z=3.7$, we see a largely similar structure. The halo is a bit brighter, mostly due to the strong $(1+z)^4$ dependence of surface brightness dimming. We can also see the bright structure of the merging massive halo at the bottom of the image. It is worth noting that while there are two `cavities' in Ly$\alpha$ to the left and right of the main halo, these are due to the structure of accreting gas, and lack of cold gas in these parts of the halo, rather than an effect of ionisation.

In the middle of the merger, a further $\sim\!260$\,Myr later, we see the central, bright part of the Ly$\alpha$ halo is significantly larger. This is unsurprising given the large amount of cold gas throughout the halo during the merger, visible in Fig.~\ref{fig:maps}. A field of view only covering the central part of the halo, or shallower observations, may therefore see a map exhibiting approximate spherical geometry, though a wider view still shows signatures of filamentary inflow. By the final snapshot, another $\sim\!440$\,Myr later, the halo no longer shows clear signs of filamentary structure. Cold gas is much less able to penetrate from outside the hot halo, leaving only bright clumps associated with galaxies themselves. 

The intrinsic emission, in the right-hand column of Fig.~\ref{fig:Lyamaps}, has two main differences with respect to the observed emission. In the centre, intrinsic emission is much brighter. We find this is due to the attenuation of emission by dust in the observed emission, which we explore further in subsequent sections. The lack of Ly$\alpha$ resonant scattering in the intrinsic output also means the emission significantly decreases at smaller radii---resonant scattering increases the observed size of Ly$\alpha$ haloes.

\subsection{Radial profiles of Ly$\alpha$ emission}

We show azimuthally averaged radial profiles of Ly$\alpha$ emission in Fig.~\ref{fig:LyaProfilesNoAGN}, with each line representing the median profile across three projections along the axes of the simulation box. Solid lines show our high resolution, shock refined simulation results, and dashed lines show the fiducial \textsc{Fable} runs. In the left-hand panel we show the observed profiles computed with \textsc{colt}, including both dust attenuation (with an assumed dust-to-metal ratio of 0.1) and resonant scattering. The profiles in this case decline fairly smoothly from a peak in the centre. Notably, the cosmological dimming corrected profile at $z=2.7$ lies below that of $z=3.3$, showing how the cold gas destruction in the halo discussed earlier begins to lead to a potentially observable effect. While the outer parts of the observed Ly$\alpha$ haloes show a reasonable match to our simulation predictions, it is clear that the surface brightness is significantly underpredicted in the halo centre, within $\sim\!50$\,kpc.

To explore the cause of the dim centre of our simulated Ly$\alpha$ haloes in more detail, in the right panel we show the \textit{intrinsic} Ly$\alpha$ emission, without attenuation or scattering. Here we can see a much better match to the observed surface brightness in the central regions of the halo, indicating that our fiducial modelling likely underestimates the efficiency with which ionising radiation is converted into Ly$\alpha$ emission in the nuclear region, or overestimates the effective attenuation by dust in the unresolved ISM. We discuss the implications of this further in Section~\ref{Section:Discussion}, and separate the effect of resonant scattering using H$\alpha$ in Section~\ref{Section:Halpha}, but this potentially points towards unresolved structure in the nuclear ISM that affects both the escape of Ly$\alpha$ photons and the local conversion of ionising photons into Ly$\alpha$ emission. These could arise from AGN-driven feedback that alters the structure of the nuclear gas, or from a better resolved, clumpier ISM that increases recombination rates and creates low-density escape channels.

We also note that the removal of dust obscuration leads to the intrinsic Ly$\alpha$ profile being less smooth than the `observed' one, because it is more dominated by individual galaxies and clumps within the halo. It is also likely that radial bins in observations are larger, which could contribute to the smoothness of the observed Ly$\alpha$ profiles. At $z=3.3$ our intrinsic emission is brighter than even the brightest observed Ly$\alpha$ halo in the QSOMuseum sample---this could be indicative of a need for obscuration, but may also be due to the enhanced cool gas mass present at that time due to the ongoing merger. A larger simulated sample would be needed to investigate this further. The intrinsic haloes also drop off considerably faster than the observed ones---as we show below, the impact of resonant scattering is particularly important at large radii. 

\subsubsection{CGM resolution effects} \label{Section:Resolution}

The dashed lines in Fig.~\ref{fig:LyaProfilesNoAGN} show the Ly$\alpha$ surface brightness profiles for the fiducial \textsc{Fable} simulation, which has considerably lower resolution in the CGM than our shock refined runs. In the observed profiles in the left panel we see how the Ly$\alpha$ haloes are considerably dimmer at lower resolution, where the median of the surface brightness increase in the high resolution run across all redshifts is a factor of 2.8. There is a smaller difference in the intrinsic Ly$\alpha$ profile (without attenuation or scattering), with a median increase of a factor of 1.9, likely reflecting the fact the ISM---which dominates the emission in this case---is at the same resolution in both runs. 

These trends are consistent with the reduced amount of cold CGM gas typically found in lower-resolution simulations \citep[see e.g.][]{vandeVoort2019,Bennett2020}: higher-resolution CGM modelling allows more multiphase, dense structures to form, which in turn boosts the Ly$\alpha$ emission. It is a strength of the present work that we can explicitly demonstrate how sensitive Ly$\alpha$ observables are to CGM resolution, and we have verified the cool gas mass is notably increased in our simulation compared to the fiducial \textsc{Fable} simulation. However, it also underlines a broader challenge to the field, as realistic predictions require simulations that resolve both the internal structure of galaxies \textit{and} the multiphase CGM.

\subsubsection{The impact of dust attenuation, resonant scattering, and AGN ionisation} \label{Section:Variations}

\begin{figure}
    \centering
    \includegraphics[width=1\linewidth]{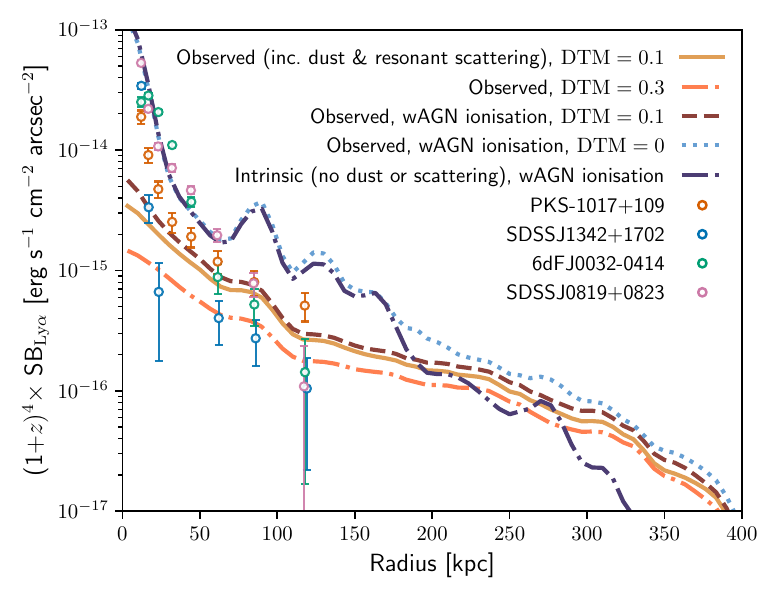}
    \caption{Radial profiles of Ly$\alpha$ surface brightness for different models at $z=3.3$, corrected for the effect of cosmological dimming. Coloured points show observational data for a few of the largest haloes in the QSOMuseum sample \citep{ArrigoniBattaia2019,GonzalezLobos2025}. We include variations in the ionisation sources used (with and without AGN), the dust content (with DTM ratios of 0, 0.1, and 0.3), and the inclusion of scattering. AGN ionisation makes little difference to the profiles, but both dust attenuation and resonant scattering can lead to notable changes in the centre and outside of the halo, respectively.}
    \label{fig:LyaProfileVariations}
\end{figure}

We have seen that considering intrinsic Ly$\alpha$ emission gives a better match to observed profiles in the inner regions of Ly$\alpha$ haloes. In this section we further explore the impact of changing our model assumptions. In Fig.~\ref{fig:LyaProfileVariations} we show surface brightness profiles for the same snapshot at $z=3.3$, with five different model choices. The trends we show at this redshift are illustrative and similar at all other times we have investigated. The solid, light orange line shows the `observed' profile reproduced from the left-hand panel of Fig.~\ref{fig:LyaProfilesNoAGN}, for reference, and does not include the effect of AGN ionisation.

We first show a model with a higher DTM ratio of 0.3 and no AGN ionisation. This value is closer to that inferred for the Milky Way and is therefore a useful comparison given the central galaxy's metallicity. The increased dust abundance further suppresses the Ly$\alpha$ surface brightness, particularly within the central regions of the halo. This demonstrates that the normalisation of the central Ly$\alpha$ profile is highly sensitive to the assumed dust model, while the qualitative behaviour of the extended halo remains unchanged.

The next variation we consider is the inclusion of an AGN as an ionising source. The dark orange dashed line in Fig.~\ref{fig:LyaProfileVariations} shows this profile, which lies very close to, if slightly above, the run without AGN ionisation. The small difference between these profiles suggests that the halo is already highly ionised by stellar radiation, so the total ionising photon budget is not the primary limitation for producing Ly$\alpha$ emission. Our halo is already quite massive by $z=3$, with a substantial hot halo already in place, and is actively forming stars---such objects may not need an actively accreting black hole in order to possess extended Ly$\alpha$ haloes. This is further motivated by the SB profile observed around the faint quasar sample in \citet{GonzalezLobos2025}, which has a similar central brightness to our `observed' profile, although the simulated Ly$\alpha$ haloes remain more extended. Even in the case where AGN ionisation only makes a small contribution, feedback from the AGN is likely required to create channels for ionising photons to escape the ISM. 

We show the intrinsic Ly$\alpha$ emission from the halo, with AGN ionisation active, as the dark blue dot-dashed line. As a reminder, the intrinsic output does not include dust attenuation, dust scattering or Ly$\alpha$ resonant scattering. As in the right-hand panel of Fig.~\ref{fig:LyaProfilesNoAGN}, we see the inner part of the halo matches the observed surface brightness profiles much better primarily due to the absence of dust attenuation, although unresolved nuclear gas structure may also affect the local production and escape of Ly$\alpha$ photons. At intermediate radii, between $\sim\!50$ and $\sim\!120$\,kpc, the intrinsic Ly$\alpha$ emission lies above observations, though as previously discussed this is likely a physical effect due to enhanced amounts of cool gas due to the ongoing merger. 

On the outskirts of the protocluster, however, the intrinsic Ly$\alpha$ halo truncates at a smaller radius than any of the runs including Ly$\alpha$ resonant scattering. We investigate this further in the final variation, including the resonant scattering of Ly$\alpha$ photons but assuming zero dust attenuation, shown in the light blue dotted line. In this variation, the central 100\,kpc is essentially identical to the intrinsic emission, pointing to the importance of attenuation in the central galaxy. Ly$\alpha$ haloes have been difficult to detect around massive, dusty galaxies, further supporting this point \citep{GonzalezLobos2023}. However, outside of $\sim\!170$\,kpc, the Ly$\alpha$ halo continues significantly further than the intrinsic emission due to the effect of Ly$\alpha$ resonant scattering. While this distinction is below the detection limits of existing observations, future surveys focussing on wider areas around protoclusters may be able to probe this regime and infer the impact of scattering.

\subsection{Emission covering fraction}

\begin{figure}
    \centering
    \includegraphics[width=1\linewidth]{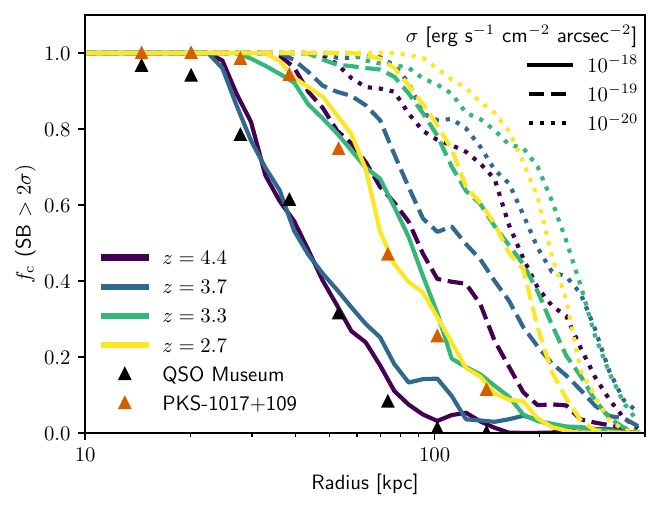}
    \caption{Covering fraction of intrinsic Ly$\alpha$ emission with S/N $> 2$, for SB limits $\sigma = 10^{-18}$, $10^{-19}$, and $10^{-20}$\,erg\,s$^{-1}$\,cm$^{-2}$\,arcsec$^{-2}$. We show the average value for three orthogonal projections along the box axes. We compare to lower limits from observational data from the full QSOMuseum sample (black triangles) and from the ELAN PKS-1017+109 (orange triangles), both from \citet{ArrigoniBattaia2019}, which have a SB limit close to $10^{-18}$\,erg\,s$^{-1}$\,cm$^{-2}$\,arcsec$^{-2}$.}
    \label{fig:LyaCoverFrac}
\end{figure}

In Fig.~\ref{fig:LyaCoverFrac} we show the covering fraction of intrinsic Ly$\alpha$ emission with SB above 2$\sigma$, where we vary the SB limit $\sigma$ between $10^{-18}$ and $10^{-20}$\,erg\,s$^{-1}$\,cm$^{-2}$\,arcsec$^{-2}$. Note that here we do not adjust the simulated maps to take account of cosmological dimming, such that this a direct prediction of observability for a given SB limit. We show the intrinsic emission as this better matches the observational data in the centre of the halo; using the `observed' data shrinks the Ly$\alpha$ haloes at the lowest sensitivity due to dust attenuation, and increases the size of the haloes at the highest sensitivity because of resonant scattering.

Fig.~\ref{fig:LyaCoverFrac} shows that deeper data leads to larger halo sizes as the fainter outskirts of the Ly$\alpha$ halo become detected. Higher sensitivity also leads to smaller differences with redshift---the brightest parts of the halo seem to vary most in our simulations as the halo evolves, though as discussed earlier this could be partly due to simplistic ISM modelling. 

Focussing on a SB limit of $10^{-18}$\,erg\,s$^{-1}$\,cm$^{-2}$\,arcsec$^{-2}$ (solid lines) we see a clear redshift divide, with the halo becoming more extended at $z \lesssim 3.3$, after a major merger heats the halo and increases the total halo mass. Interestingly, these Ly$\alpha$ halo sizes in the simulation correspond closely to the two observed profiles we show, that have a similar SB limit of $\sim\!10^{-18}$\,erg\,s$^{-1}$\,cm$^{-2}$\,arcsec$^{-2}$ \citep{ArrigoniBattaia2019}. The first set of observational data shows the average covering fraction across the QSOMuseum sample (black triangles), which aligns well with the higher redshift snapshots of our simulation. The second, showing the ELAN PKS-1017+109 (orange triangles), lies close to our lower redshift snapshots. This potentially implies an evolution in Ly$\alpha$ haloes---with a more compact halo representative of the wider population evolving into an ELAN-like object during a merger phase, and after a significant episode of halo heating---however we need a larger sample of high-resolution cluster simulations to investigate this further.

\subsection{Comparison to H$\alpha$} \label{Section:Halpha}

\begin{figure}
    \centering
    \includegraphics[width=\linewidth]{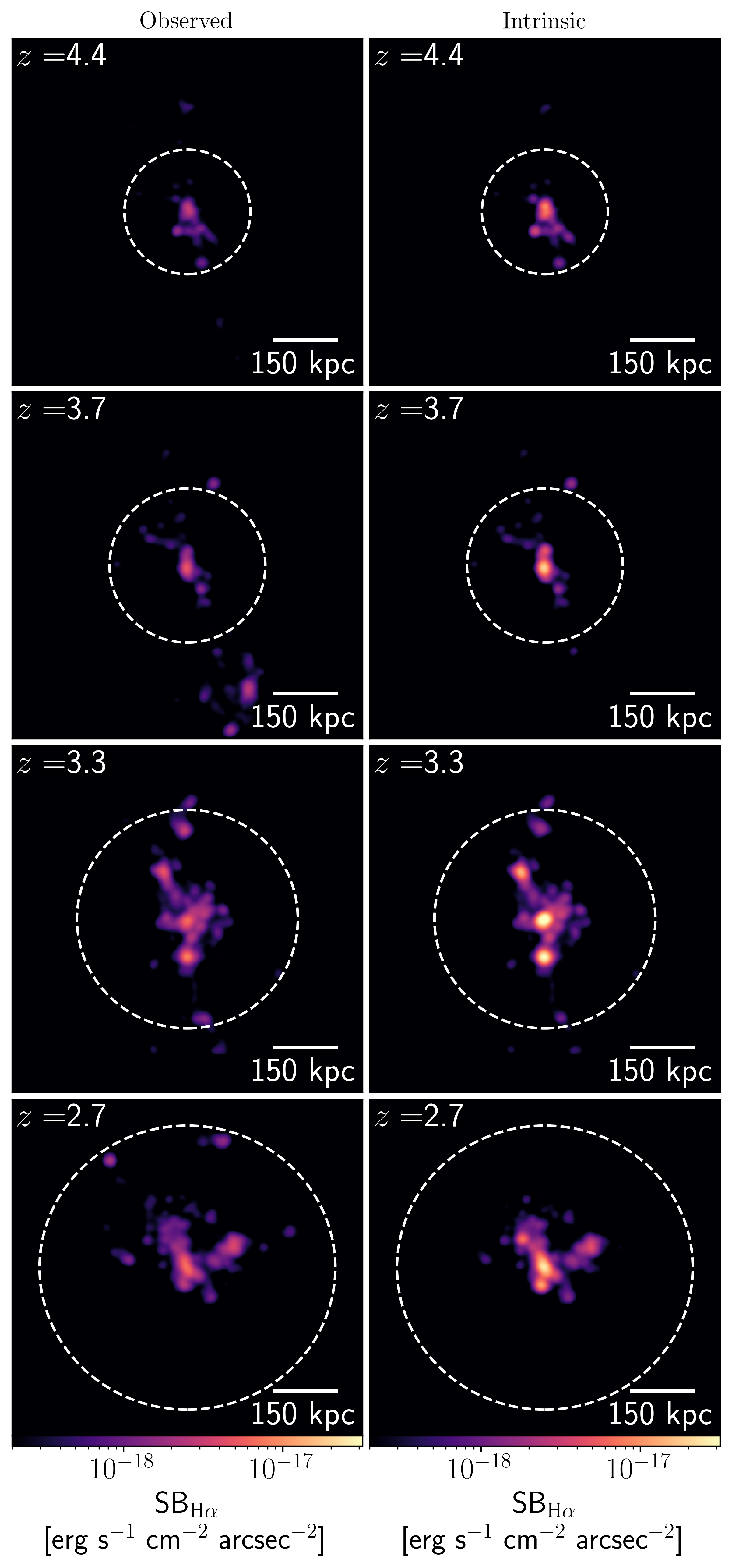}
    \caption{Maps of H$\alpha$ surface brightness (with no AGN ionisation), convolved with a 1 arcsecond Gaussian, at four redshifts $z=4.4,3.7,3.3,$ and $2.7$ (from top to bottom). The projections are the same shown in Figs.~\ref{fig:maps} and \ref{fig:Lyamaps}. The left column shows observed maps, including the effects of attenuation and scattering by dust, and the right column shows intrinsic maps, showing pure emission without these effects. Dashed white circles in each panel show $R_{200}$. Like Ly$\alpha$, H$\alpha$ traces the presence of dense, neutral gas, however H$\alpha$ emission is much dimmer and less extended.}
    \label{fig:Hamaps}
\end{figure}

\begin{figure}
    \centering
    \includegraphics[width=1\linewidth]{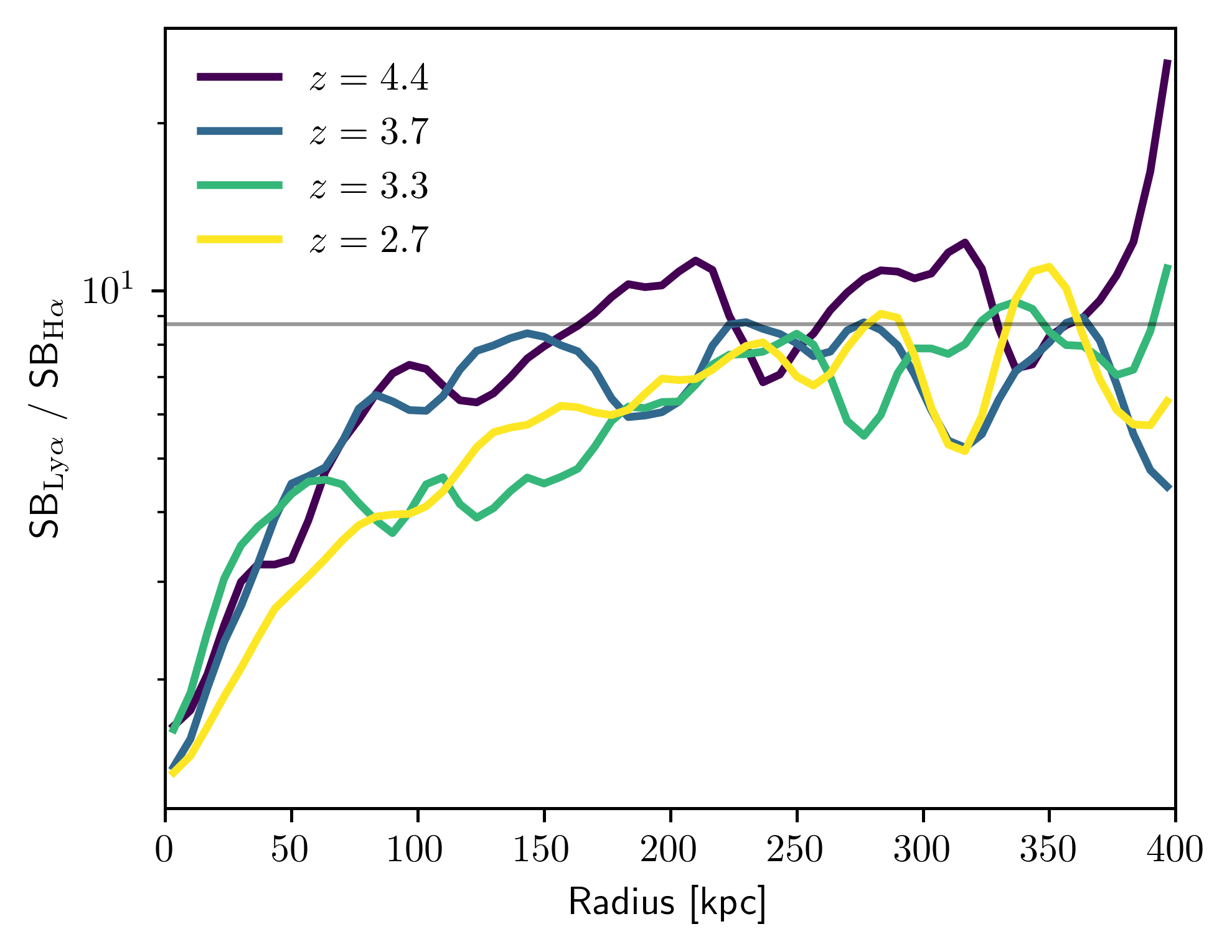}
    \caption{Profiles of the surface brightness ratio of Ly$\alpha$ to H$\alpha$ at four redshifts, for the `observed' profiles calculated from our simulations. The horizontal grey line shows the canonical case B recombination reference value of 8.7 \citep{Leibler2018}. Ly$\alpha$ is significantly brighter than H$\alpha$ at all radii studied. The lower ratio within $\sim\!50$\,kpc, which is consistent with existing observations, is caused by dust attenuation.}
    \label{fig:LyaHaRatio}
\end{figure}

Ly$\alpha$ is a resonant line that has several possible emission mechanisms, complicating our ability to infer CGM properties. A potential solution that is currently being investigated is by instead using the H$\alpha$ line, which does not suffer from the same complications. While H$\alpha$ is shifted into the IR at $z\gtrsim2$, making it hard to observe from the ground, in the era of \textit{JWST} mapping H$\alpha$ becomes a feasible prospect. In this section, we therefore also make predictions for the brightnesses and extent of H$\alpha$ haloes. As discussed earlier, we will explore predictions for kinematics in future work. 

Fig.~\ref{fig:Hamaps} shows maps of H$\alpha$, for the same snapshots and orientations as the Ly$\alpha$ maps in Fig.~\ref{fig:Lyamaps}. We also choose our colour map limits to be the same as for Ly$\alpha$, highlighting that H$\alpha$ is considerably dimmer. We predict that H$\alpha$ haloes are much more compact than Ly$\alpha$ at fixed surface brightness. Bright spots in the image are associated with galaxies themselves, with a much smaller contribution from extragalactic gas and inflowing filaments. The intrinsic H$\alpha$ emission (right column), as with Ly$\alpha$, is brighter in the centre due to a lack of obscuration. 

In Fig.~\ref{fig:LyaHaRatio} we show the ratio of `observed' Ly$\alpha$ to H$\alpha$ surface brightness profiles. In the central $\sim\!50$\,kpc of the haloes we find an average ratio of 3--5, consistent with the available observations at slightly lower redshift \citep{Leibler2018,Langen2023}. The ratio increases with radius until flattening around the case B reference value of 8.7 \citep[][]{Leibler2018}, shown as a horizontal grey line. In the intrinsic emission, the ratio remains approximately at the case B reference value throughout the centre of the halo, pointing towards dust attenuation as the cause of decline in ratio at low radii.

\section{Discussion} \label{Section:Discussion}

We first compare our results to some existing simulation work, before discussing the caveats of our analysis. 

\subsection{Cold gas content}

The evolution of the ICM in the TNG-Cluster simulations \citep{Nelson2024} was explored by \citet{Rohr2025}. TNG-Cluster has a sample of 352 $z=0$ galaxy clusters and their progenitors, a much larger sample than ours, though its resolution is considerably coarser---cells are more massive by a factor of $\sim\!350$. They draw largely similar conclusions to ours, with a multiphase halo fed by filamentary accretion at $z=4$ transitioning by $z=2$ to a hot ICM with cold gas primarily restricted to (and sourced by) satellite galaxies. This happens across the cluster halo mass range, with the proto-ICM being cooler at high redshifts. The transition they find also appears to be driven by AGN feedback from the inside out, in agreement with our conclusions. Our case-study shows how the CGM to ICM transition can happen rapidly, potentially accelerated by merger activity. 

In our Fig.~\ref{fig:MgIIHist} we highlight how the highest column densities of Mg\,\textsc{ii}, associated with satellites, are the most able to survive the CGM to ICM transition. The quantity of cold material in the ICM at lower redshift is therefore dominated by the gas that sources these high columns. We therefore support the conclusions of \citet{Rohr2025}, who find the cold ICM gas content correlates with the number of gaseous satellites. 

\subsection{Ly$\alpha$ haloes}

Predictions for Ly$\alpha$ haloes in the TNG50 simulations \citep{TNG50_1,TNG50_2} were made by \citet{Byrohl2021}. TNG50 also has high resolution throughout the simulation (gas cells are an average factor of $\sim\!2.5$ more massive than our simulation) and the same ISM model. Given the size of the cosmological volume, however, they focus on lower mass haloes, mostly $M_\mathrm{h} \lesssim 10^{12}\,\mathrm{M}_\odot$. Both cluster-scale shock heating and AGN feedback are therefore likely to play a larger role in our halo. They also use a different model for Ly$\alpha$ photon sourcing, with emissivity linked to the star formation rate of gas cells instead of being sourced from star particles themselves. Despite these differences, however, it is worth comparing their key results.

\citet{Byrohl2021} find a flattening in Ly$\alpha$ halo profiles at large radii which they attribute to photon contributions from neighbouring haloes. We do find the SB profile somewhat flattens when Ly$\alpha$ resonant scattering is included, though the SB continues to decrease at a shallower rate. This is likely to be a consequence of the radial cut we used in our Ly$\alpha$ calculation---we do not have photons from outside this region to illuminate the halo outskirts. Whereas \citet{Byrohl2021} find resonant scattering of star-formation driven Ly$\alpha$ photons sets the inner brightness of Ly$\alpha$ haloes, our intrinsic versus attenuated comparison (Fig.~\ref{fig:LyaProfileVariations}) suggests that---at least in protoclusters---it is the ability of photons to \textit{escape} the ISM that sets the central SB. 

This interpretation is supported by \citet{Costa2022}, whose radiative hydrodynamical zooms of a $z=6$ quasar host showed that AGN feedback is required for Ly$\alpha$ photons to escape the central galaxy (whether those photons are sourced from stars or the black hole itself). They additionally test a model with Ly$\alpha$ photons emitted from the broad line region (BLR) of the quasar itself, which even by itself can reproduce observed $z\sim6$ SB profiles. Potentially the addition of a BLR source into our model could bring the central SB of our `observed' profiles into better agreement with observations, though we note this is still degenerate with the modelling of Ly$\alpha$ escape from the ISM.

\subsection{Caveats to our analysis}

The simplistic modelling of the ISM and the ISM--CGM boundary in our simulation is a key limitation of this work. We have shown how the SB of Ly$\alpha$ can vary significantly when our assumptions about photon escape from the galaxy are changed. However, we can conclude from our work that the inner parts of Ly$\alpha$ nebulae can only be powered by stellar emission when almost all ionising photons are allowed to escape the galaxy. We therefore require either a lower effective attenuation than predicted by a Milky-Way-like dust model, a significant contribution from AGN-powered Ly$\alpha$ emission, or unresolved ISM structure that enhances Ly$\alpha$ escape and production efficiencies. In future work, including an additional central source of Ly$\alpha$ photons would allow us to investigate this further, though a full understanding may only be accessible by simulating a similar object with a better resolved multi-phase ISM model \citep{Hopkins2018,Marinacci2019,Kannan2025}, potentially also in combination with additional CGM refinement, though this would be an extremely expensive simulation for such a massive object.

The version of the refinement scheme used in this work means we only have a single black hole in the simulation (see Section~\ref{Section:Methods:Simulations}). While we have verified the galaxy properties are comparable to the fiducial \textsc{Fable} simulation, this means we do not capture the impact of dual or multiple quasars on the illumination of Ly$\alpha$ haloes in this work. With future simulations, containing multiple active black holes, we will then be able to compare to observations of intergalactic bridges connecting quasar pairs \citep{ArrigoniBattaia2019b,Herwig2024}.

In this work we also only focus on an individual object, due to the computational expense of maintaining a high resolution CGM for a massive object. Our case study has highlighted a number of follow-up questions, though further exploration of, for example, the lifetimes and host halo masses of ELAN-like objects will require a larger sample of haloes across a wider range in mass, also performed to lower redshift. 

\section{Conclusions} \label{Section:Conclusions}

While galaxy clusters at low redshift have been extensively studied, their early growth and evolution is much less certain. The key transition here is from galaxies closely interacting with a multiphase CGM to galaxies being embedded in a hot, strongly X-ray emitting ICM. In this paper we present simulations of a protocluster passing through this transition, with enhanced numerical resolution allowing us to probe its gaseous halo in detail. We present both properties directly from the simulation and mock observables to study ``the birth of the ICM'', and our key conclusions are as follows:

\begin{enumerate}
    \item The protocluster regime is dominated by multiphase gas, with cold filaments supplying significant material into the halo. At lower redshifts, as the halo is heated by feedback, much of this multiphase CGM is destroyed.
    \item The gaseous halo becomes hotter, more smoothly distributed, and more uniformly enriched (to the canonical $0.3$\,Z$_\odot$) during this process, with the volume-filling phase entering the X-ray regime and becoming much closer to the observed lower redshift ICM. 
    \item Intermediate column densities of cold-phase tracers like Mg\,\textsc{ii} are preferentially destroyed in this process. The densest gas associated with satellites however survives this transition, leading to a much clumpier distribution as the halo evolves.
    \item The CGM to ICM transition happens from the inside out. Cool gas is destroyed and becomes further ionised in the inner halo (at radii $\sim\!50-250$\,kpc) first, and material is redistributed to larger radii.
    \item Extended Ly$\alpha$ emission is present at all redshifts we studied, and remains detectable even without AGN ionisation. The halo morphology shifts from filamentary to more spherical as it evolves.
    \item Assuming a simple dust model leads to substantial uncertainty in the central Ly$\alpha$ surface brightness. Our DTM = 0.1 model already underpredicts the central emission of the Ly$\alpha$ haloes, while a higher DTM = 0.3 model suppresses the emission even further. This suggests that the effective Ly$\alpha$ production and escape in the nuclear region are more efficient than captured by our model, likely reflecting unresolved ISM structure and/or additional AGN-powered Ly$\alpha$ emission.
    \item Turning on AGN photoionisation produces only a modest change in the extended Ly$\alpha$ SB profile, consistent with the halo already being highly ionised by stars in this massive system. This suggests that the total ionising photon budget is not the primary limitation for producing Ly$\alpha$ emission in the halo. However, feedback is very likely needed to allow sufficient numbers of ionising photons to escape from the central galaxy \citep[see also][]{Costa2022}. Resonant scattering of Ly$\alpha$ photons only seems to impact halo outskirts, at larger radii than what is currently detectable.
    \item Increased numerical resolution tends to boost inferred SB profiles, due to increased amount of cold gas at higher resolution, which may be an issue for many existing simulations of galaxy clusters.
    \item H$\alpha$ haloes are considerably smaller and dimmer than Ly$\alpha$ haloes, but may be easier to interpret, which is particularly exciting in the era of \textit{JWST}. The central Ly$\alpha$/H$\alpha$ brightness ratio is consistent with current measurements.
    \item The evolution of Ly$\alpha$ covering fractions at fixed SB is consistent with a scenario where a `typical' nebula can transition to an ELAN-like state following a major heating episode, but confirming this will require a larger suite of similarly resolved protocluster simulations.
\end{enumerate}

The case study presented here suggests that the emergence of the early ICM at $z\sim3$ can be thought of as a coupled transition: the disappearance of intermediate-column cold absorbers, the growth and subsequent over-ionisation of hot-phase tracers in the inner halo, and a morphological shift in Ly$\alpha$ emission from filamentary inflow to a smoother, clump-dominated nebula. During this transition the profiles of density, temperature and metallicity all flatten as material is redistributed outward. Intriguingly, this occurs near a halo mass of $M_\mathrm{h}\sim\!10^{13}$\,M$_\odot$, where baryon redistribution is expected to be maximally efficient. 

To explore this transition further requires a larger sample of highly resolved, simulated protoclusters. Addressing the largest uncertainties in this work will also need improved sub-grid models for the ISM, which will allow self-consistent predictions for both absorption and emission of gaseous haloes. Ongoing and upcoming observations across the electromagnetic spectrum will detect line emission with MUSE and \textit{JWST}, X-ray signatures of hot phases with \textit{ATHENA}, and even SZ detections of individual haloes before cosmic noon. Combining these multi-wavelength data with a new suite of cluster simulations will offer a promising path towards fully understanding the birth of the ICM in forming galaxy clusters.

\section*{Acknowledgments}

We thank the referee for a constructive report which improved the manuscript. JSB acknowledges support from the Simons Collaboration on Learning the Universe and a Leverhulme Trust Early Career Fellowship. D.S. acknowledges support from the Science and Technology Facilities Council (STFC) under grant ST/W000997/1. Support for C.L. was provided by NASA through the NASA Hubble Fellowship grant \#HST-HF2-51538.001-A awarded by the Space Telescope Science Institute, which is operated by the Association of Universities for Research in Astronomy, Inc., for NASA, under contract NAS5-26555. The simulations presented in this work were performed using the Cambridge Service for Data Driven Discovery (CSD3) operated by the University of Cambridge Research Computing Service (\href{https://www.csd3.cam.ac.uk}{www.csd3.cam.ac.uk}), provided by Dell EMC and Intel using Tier-2 funding from the Engineering and Physical Sciences Research Council (capital grant EP/P020259/1). This work made use of the NumPy \citep{Numpy}, SciPy \citep{SciPy}, and Matplotlib \citep{Matplotlib} \textsc{python} packages.

\bibliographystyle{aasjournal}

\bibliography{main}

\end{document}